\documentclass[journal]{IEEEtran}

\usepackage{amssymb}
\usepackage[cmex10]{amsmath}
\usepackage{stfloats}
\usepackage{graphicx}
\usepackage{subfigure}
\usepackage{tabularx}
\usepackage{epsfig,epsf,color,balance,cite}
\usepackage{verbatim}
\usepackage{url}
\usepackage{bm}

\newtheorem{theorem}{Theorem}
\newtheorem{lemma}{Lemma}

\usepackage{algorithm}
\usepackage{algorithmic}

\begin{document}

% paper title
\title{Energy Efficient Resource Allocation in Machine-to-Machine Communications with Multiple Access and Energy Harvesting for IoT}

\author{
\IEEEauthorblockN{Zhaohui Yang,
                  Wei Xu, \IEEEmembership{Senior Member, IEEE},
                  Yijin Pan,
                  Cunhua Pan, and
                  Ming Chen
                  }

\thanks{This work was supported in part by the National Nature Science Foundation of China under grants 61471114, 61372106 \& 61221002, in part by the Six Talent Peaks project in Jiangsu Province under grant GDZB-005,  in part by the UK Engineering and Physical Sciences Research Council under Grant EP/N029666/1, and in part by the Scientific Research Foundation of Graduate School of Southeast University under Grant YBJJ1650. This paper was presented at the IEEE Infocom Workshops 2017 in Atlanta, GA, USA \cite{Yang2017EnergyM2M}. (\emph{Corresponding authors: Wei Xu; Cunhua Pan}.)}

\thanks{Z. Yang, W. Xu, Y. Pan, and M. Chen are with the National Mobile Communications Research
Laboratory, Southeast University, Nanjing 210096, China, (Email: \{yangzhaohui, wxu, panyijin, chenming\}@seu.edu.cn).}
 \thanks{C. Pan is with the School of Electronic Engineering and Computer Science, Queen Mary, University of London, London E1 4NS, U.K., (Email: c.pan@qmul.ac.uk).}
}

\maketitle

\begin{abstract}
This paper studies energy efficient resource allocation for a machine-to-machine (M2M) enabled cellular network with non-linear energy harvesting, especially focusing on two different multiple access strategies, namely non-orthogonal multiple access (NOMA) and time division multiple access (TDMA). Our goal is to minimize the total energy consumption of the network via joint power control and time allocation while taking into account  circuit power consumption. For both NOMA and TDMA strategies, we show that it is optimal for each machine type communication device (MTCD) to transmit with the minimum throughput, and the energy consumption of each MTCD is a convex function with respect to the allocated transmission time. Based on the derived optimal conditions for the transmission power of MTCDs, we transform the original optimization problem for NOMA to an equivalent problem which can be solved suboptimally via an iterative power control and time allocation algorithm. Through an appropriate variable transformation, we also transform the original optimization problem for TDMA to an equivalent tractable problem, which can be iteratively solved. Numerical results verify the theoretical findings and demonstrate that NOMA consumes less total energy than TDMA at low circuit power regime of MTCDs, while at high circuit power regime of MTCDs TDMA achieves better network energy efficiency than NOMA.
\end{abstract}

\IEEEpeerreviewmaketitle

\section{Introduction}

Machine-to-machine (M2M) communications have been considered as one of the promising technologies to realize the Internet of Things (IoT) in the future 5th generation network.
M2M communications can be applied to many IoT applications, which mainly involve new business models and opportunities, such as smart grids, environmental monitoring and intelligent transport systems \cite{Wu2011M2M}.
Different from conventional human type communications, M2M communications have many unique features \cite{Zheng2012Radio}.
The unique features include massive transmissions from a large number of machine type communication devices (MTCDs), small bursty natured traffic (periodically generated), extra low power consumption of MTCDs, high requirements of energy efficiency and security.

A key challenge for M2M communications is access control, which manages the engagement of massive MTCDs to the core network.
To tackle this challenge, various solutions have been proposed, e.g., by using wired access (cable, DSL) \cite{6525600}, wireless short distance techniques (WLAN, Bluetooth), and wide area cellular network infrastructure (Long Term Evolution-Advanced (LTE-A), WiMAX)\cite{6845044}.
%**** sth more specfic *********
Among all these solutions, an effective approach is to deploy machine type communication gateways (MTCGs) to act as relays of MTCDs \cite{Zheng2012Radio}.
With the help of MTCGs, all MTCDs can be successfully connected to the base station (BS) at the additional cost of energy consumption \cite{6786066,7010526,6824817,7420617}.
To enable multiple MTCDs to transmit data to the same MTCG, time division multiple access (TDMA) was adopted in \cite{Zhang_2016}.
However, since there are a vast number of MTCDs to be served, TDMA leads to large transmission delay and synchronization overhead.
By splitting users in the power domain, non-orthogonal multiple access (NOMA) can simultaneously serve multiple users at the same frequency or time resource \cite{saito2013non}.
Consequently, NOMA based access scheme yields a significant gain in spectral efficiency over the conventional orthogonal TDMA \cite{saito2013non,Yang2017MultiCell,ding2017application,Chen2014ULNOMA,Imari2014ULNOMA,Zhang2016ULNOMA}.
This favorable characteristic makes NOMA an attractive solution for supporting massive MTCDs in M2M networks.
Considering NOMA, \cite{Yang2017Energy} investigated an M2M enabled cellular network, where multiple MTCDs simultaneously transmit data to the same MTCG and multiple MTCGs simultaneously transmit the gathered data to the BS.

Besides, another key challenge is the energy consumption of MTCDs \cite{7378918,6644241,7263374}.
According to \cite{Zheng2012Radio}, the total system throughput of an M2M network is mainly limited by the energy budget of the MTCDs.
To improve the system performance, energy harvesting (EH) has been applied to wireless communication networks \cite{4595260,Bi2015Wireless,Huang2015Energy,7001217}.
In particular,  direct and non-direct energy transfer based schemes for EH were investigated in [23], while in [24], the optimization of green-energy-powered cognitive radio networks was surveyed.
Recently, the downlink resource allocation for EH in small cells was studied in \cite{7317521,7562484,7948714}.
By using EH, MTCDs are able to harvest wireless energy from radio frequency (RF) signals \cite{6477832,7944540,7555321,Yang2017EnergyM2MEH}, and the system energy can be significantly improved.
Consequently, implementing EH is promising in M2M communications especially with MTCDs configured with low power consumption.
In previous wireless powered communication networks using relays \cite{6552840,7342973,7470278}, it was assumed that an energy constrained relay node harvests energy from RF signals and the relay uses that harvested energy to forward source information to destination.
Due to the extra low power budget of MTCDs in M2M communications, it is reasonable to let the MTCD transmit data to an MTCG, and then the MTCG relays the information while the MTCD simultaneously harvests energy from the MTCG, which is different from existing works, e.g., \cite{6552840,7342973,7470278}.
Enabling the source node to harvest energy from the relay node, a power-allocation scheme for a decode-and-forward relaying-enhanced wireless system was proposed in \cite{Huang2016Optical} with one source node, one relay node and one destination node.

The above-mentioned energy consumption models, considered in \cite{Yang2017EnergyM2M,6824817,7420617,Zhang_2016,Yang2017Energy,7378918,6644241,6477832,7944540,7555321,Yang2017EnergyM2MEH}, are only concerned with the RF transmission power and ignore the circuit power consumption of MTCDs and MTCGs.
However, as stated in \cite{Kim2010Leveraging}, the circuit power consumption is non-negligible compared to RF transmission power.
Without considering  the circuit power consumption, energy saving can always benefit from longer transmission time \cite{angelakis2014minimum,Chin2015Power}.
While considering the circuit power consumption which definitely exists in practice, the results vary significantly in that it can not be always energy efficient for long transmission time due to the fact that the total energy consumption becomes infinity as transmission time goes infinity.
Hence, it is of importance to investigate the optimal transmission time when taking into account the circuit power consumption in applications.

Given access control and energy consumption challenges in M2M communications, both TDMA and NOMA based M2M networks with EH are proposed in this work.
The MTCDs first transmit data to the corresponding MTCGs, and then MTCGs transmit wireless information to the BS and wireless energy to the MTCDs.
To prolong the lifetime of the considered network, the harvested energy for each MTCD is set to be no less than the consumed energy in information transmission (IT) stage.
The main contributions of this paper are summarized as follows:
\begin{itemize}
  \item  We formulate the total energy minimization problem for the M2M enabled cellular network with non-linear energy harvesting (EH) model via joint power control and time allocation.
      In the non-linear EH model, we consider the receiver sensitivity, on which energy conversion starts beyond a threshold.
      Besides, we explicitly take into account the circuit energy consumption of both MTCDs and MTCGs.
      All theses factors are critical in practical applications which inevitably affect the system performance. Specifically,
      the non-linear EH model leads to a non-smooth objective function and non-smooth constraints, and the circuit energy consumption affects the optimal transmission time of the system.
  \item  For the NOMA strategy, we observe that: 1) it is optimal for each MTCD to transmit with minimal throughput; 2) it is further revealed that the energy consumption of each MTCD is a convex function with respect to the allocated transmission time.
      Given these observations, it indicates that a globally optimal transmission time always exists that the optimal transmission time equals the maximally allowed transmission time if it does not exceed a quantified threshold derived in closed form.
  \item
      To solve the original total energy minimization problem for the NOMA strategy, we devise a low-complexity iterative power control and time allocation algorithm.
      Specifically, to deal with the non-smooth EH function, we introduce new sets during which MTCDs can effectively harvest energy.
      Given new sets, the EH function of MTCDs can be presented as a continuous one.
      Moreover, to deal with nonconvex objective function, nonconvex minimal throughput constraints, and nonconvex energy causality constraints, we transform these nonconvex ones into convex ones by manipulations with the optimal conditions.
      The convergence of the iterative algorithm is strictly proved.
  \item  For the TDMA strategy, we verify that the two observations for NOMA are also valid.
      Although the original total energy minimization problem for the TDMA strategy is nonconvex, the problem can be transformed into an equivalent tractable one, which can be iteratively solved to its suboptimality.
      For the total energy minimization, numerical results identify that NOMA is superior over TDMA at small circuit power regime of MTCDs, while TDMA outperforms NOMA at large circuit power regime of MTCDs.
\end{itemize}

This paper is organized as follows.
In Section $\text{\uppercase\expandafter{\romannumeral2}}$, we introduce the system and power consumption model.
Section $\text{\uppercase\expandafter{\romannumeral3}}$ and Section $\text{\uppercase\expandafter{\romannumeral4}}$ provide the energy efficient resource allocation for NOMA and TDMA, respectively.
Numerical results are displayed in Section $\text{\uppercase\expandafter{\romannumeral5}}$
and conclusions are drawn in Section $\text{\uppercase\expandafter{\romannumeral6}}$.

\section{System and Power Consumption Model}

\subsection{System Model}
\begin{figure}
\centering
\includegraphics[width=3in]{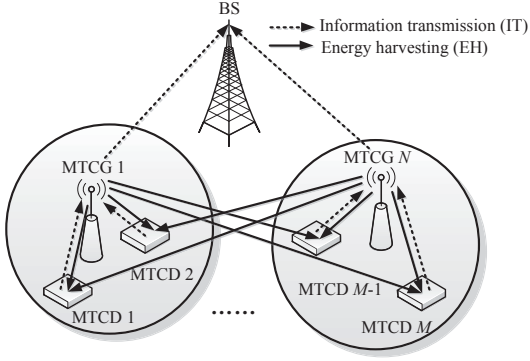}
%\vspace{-1.5em}
\caption{The considered uplink M2M enabled cellular network.}\label{sys1}
%\vspace{-1.5em}
\end{figure}

Consider an uplink M2M enabled cellular network with $N$ MTCGs and $M$ MTCDs, as shown in Fig.~\ref{sys1}.
Denote the sets of MTCGs and MTCDs by $\mathcal N=\{1, \cdots, N\}$ and $\mathcal M=\{1, \cdots, M\}$, respectively.
Each MTCG serves as a relay for some MTCDs.
Assume that the decode-and-forward protocol \cite{Karmakar2011DF} is adopted at each MTCG.
Denote ${\cal{J}}_i=\{J_{i-1}+1, \cdots, J_i\}$
as the specific set of MTCDs served by MTCG $i\in \mathcal N$, where $J_0=0$, $J_N=M$, $J_i=\sum_{l=1}^i|{\cal{J}}_l|$, and $|\cdot|$ is the cardinality of a set.\footnote{In this paper, we assume that MTCDs are already associated to MTCGs by using the cluster formation methods for M2M communications, e.g., in \cite{6503998,6649207,7558142,7913656}.
 Joint optimization of cluster formation and resource allocation in M2M communications with NOMA/TDMA and EH can certainly further improve the performance, but we leave it in future work in order to focus on the power control and time allocation in the current submission.}
To reduce the receiver complexity at the MTCG, the maximal number of MTCDs associated to one MTCG is set as four.
Obviously, we have $\bigcup_{i\in\mathcal N} \mathcal J_i=\mathcal M$.

\subsection{NOMA Strategy}

\begin{figure}
\centering
\includegraphics[width=3.35in]{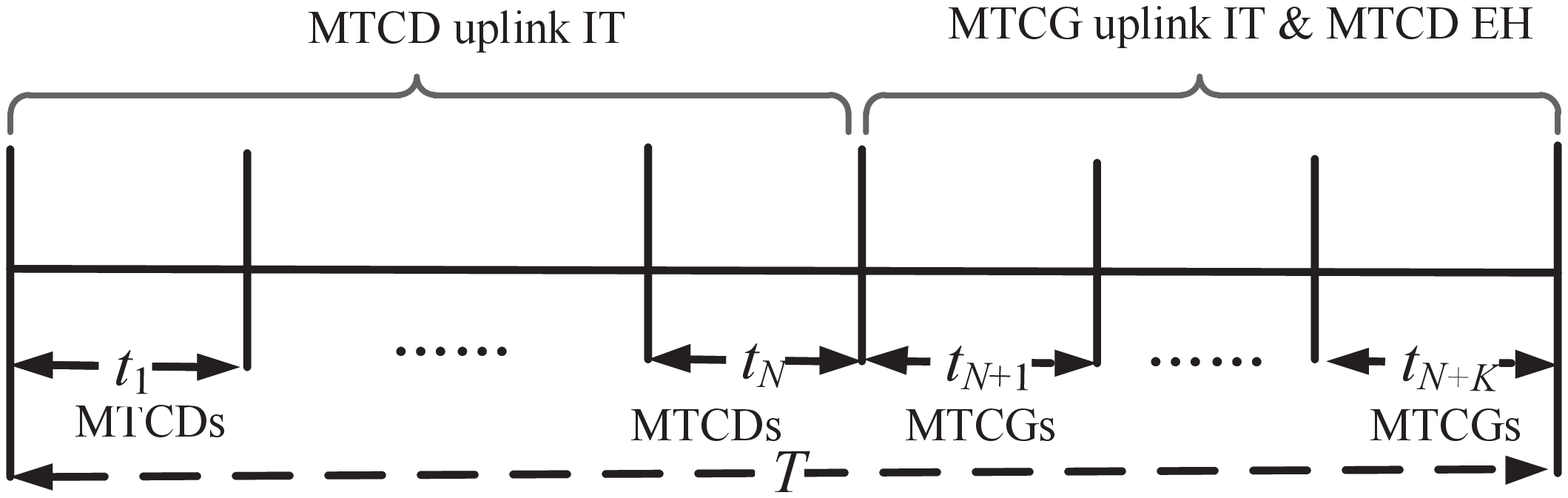}
%\vspace{-1.8em}
\caption{Time sharing scheme for NOMA strategy during one uplink transmission period.}\label{sys2}
%\vspace{-1.8em}
\end{figure}

In time constraint $T$, each MTCD  has some payloads to transmit to the BS.
By using superposition coding at the transmitter and successive interference cancellation (SIC) at the receiver, multiple MTCDs (or MTCGs) can simultaneously transmit signals to the corresponding receiver using NOMA.
To reduce the receiver complexity and error propagation due to SIC, it is reasonable for the same resource to be multiplexed by a small number (usually two to four) of devices \cite{6464495}.
Considering the receiver complexity at the BS, the sets of MTCGs are further classified into multiple small clusters.
For MTCGs, the set $\mathcal N$ is classified into $K$ clusters.
Let $\mathcal K=\{1, 2, \cdots, K\}$ be the set of clusters.
Denote ${\cal{I}}_k=\{I_{k-1}+1, \cdots, I_k\}$
as the specific set of MTCGs in cluster $k\in\mathcal K$, where $I_0=0$, $I_K=N$, $I_k=\sum_{l=1}^k|{\cal{I}}_l|$, and $|\mathcal I_k|\leq 4$.\footnote{A scheme for cluster formation for uplink NOMA is in \cite{Mohammad17uplinkNOMA}.
According to \cite{Mohammad17uplinkNOMA} and \cite{7273963}, one common scheme with two devices in each cluster is the strong-weak scheme, i.e., the device with the strongest
channel condition is paired with the device with the weakest, and
the device with the second strongest is paired with one with the
second weakest, and so on.}

Note that our major motivation of using NOMA is to enhance the ability of serving more terminals simultaneously \cite{saito2013non}.
However, the number of terminals occupying the same resource cannot be arbitrarily large in order to make NOMA effective in practice.
Therefore, if there would be an even higher number of IoT terminals, we believe that a number of ways including NOMA should be further incorporated to better address the access problem for higher number of IoT terminals \cite{Shirvanimoghaddam2017Fund}.

As depicted in Fig.~\ref{sys2}, time $T$ consists of $N+K$ uplink transmission phases for MTCDs and MTCGs.
NOMA is adopted for MTCDs to transmit data to MTCGs in the first $N$ phases.
Both NOMA and EH are in operation in the last K phases where MTCGs transmit data to the BS and simultaneously MTCDs harvest energy from all MTCGs.
In the $i$-th ($i\leq N$) phase with allocated time $t_{i}$, all MTCDs in $\mathcal J_i$ simultaneously transmit data to MTCG $i$ according to the NOMA principle and MTCG $i$ detects the signal.
In the ($N+k$)-th ($k\leq K$) phase with allocated time $t_{N+k}$, all MTCGs in $\mathcal I_k$ simultaneously transmit the decoded data from the served MTCDs to the BS by using the NOMA strategy.
As a result, we have the following transmission time constraint
\begin{equation}\label{sys1eq1}
\sum_{i =1} ^{N+K}t_i \leq T.
\end{equation}

In the $i$-th ($i\leq N$) phase, all MTCDs in $\mathcal J_i$ simultaneously transmit data to MTCG $i$ following the NOMA principle.
The received signal of MTCG $i$ is
\begin{equation}\label{sys1eq2}
y_i=\sum_{j=J_{i-1}+1}^{ J_i} h_{ij} \sqrt {p_j}s_j + n_i,
\end{equation}
where $h_{ij}$ is the channel between MTCD $j$ and MTCG $i$,
$p_j$ denotes the transmission power of MTCD $j$,
$s_j$ is the transmitted message of MTCD $j$,
and $n_i$ represents the additive zero-mean Gaussian noise with variance $\sigma^2$.
Without loss of generality, the channels are sorted as $|h_{i (J_{i-1}+1)}|^2 \geq \cdots \geq |h_{i J_i}|^2$.
By applying SIC to decode the signals \cite{Chen2014ULNOMA,Imari2014ULNOMA,Zhang2016ULNOMA}, the achievable throughput of MTCD $j\in\mathcal J_i$ is
\begin{equation}\label{sys1eq3}
r_{ij}=B t_{i}\log_2\left(
1+\frac{|h_{ij}|^2 p_j}
{ \sum_{l=j+1}^{J_i} |h_{il}|^2 p_l +\sigma^2}
\right),
\end{equation}
where $B$ is the available bandwidth for transmission. %and we define $\sum_{l=J_i+1}^{J_i} p_l=0$ for $j=J_i$.
Note that we consider the case where MTCDs associated to different MTCGs are allocated with orthogonal time resource. Therefore, the interference from other MTCDs associated to different MTCDs is ignored.

In the ($N+k$)-th ($k\leq K$) phase with allocated time $t_{N+k}$, after having successfully decoded the messages in the last $N$ phases, MTCGs in $\mathcal I_k$ simultaneously transmit the gathered data to the BS based on the NOMA principle.
Denote the channel between MTCG $i$ and the BS by $h_i$.
Without loss of generality, the channels are sorted as $|h_{I_{k-1}+1}|^2 \geq \cdots \geq |h_{I_k}|^2$, $\forall k \in \mathcal K$.
Hence, the achievable throughput of MTCG $i\in\mathcal I_k$ can be expressed as \cite{Chen2014ULNOMA,Imari2014ULNOMA,Zhang2016ULNOMA}
\begin{equation}\label{sys1eq4}
r_i=B t_{N+i}\log_2\left(
1+\frac{|h_{i}|^2 q_i}
{ \sum_{n=i+1}^{I_k} |h_{n}|^2 q_n+\sigma^2}
\right),
\end{equation}
where $q_i$ is the transmission power of MTCG $i$.

\begin{figure}
	\centering
	\includegraphics[width=3in]{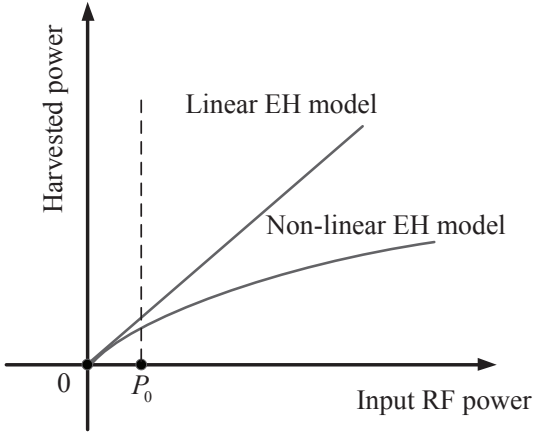}
\caption{Comparison between the linear and non-linear EH model.}\label{syseh}
\end{figure}

According to \cite{Zheng2012Radio}, MTCDs are always equipped with finite batteries, which limit the lifetime of the M2M enabled cellular network.
%Since the MTCDs are always \cite{Zheng2012Radio} equipped with limited battery, the MTCDs are active when they need to transmit data or harvest energy and inactive in other cases, i.e., the total active time for MTCD $j \in \mathcal J_i$ served by MTCG $i$ is $t_i+t_{N+1}$.
To further prolong the lifetime, EH technology is adopted for MTCDs to harvest energy remotely from RF signals radiated by MTCGs \cite{Bi2015Wireless}.
Specifically, each MTCD harvests energy when MTCGs transmit data to the BS.
Since the noise power is much smaller than the received power of MTCGs in practice \cite{Ju2014Throughput,Ju2014Optimal,zhou2014wireless},
the energy harvested from the channel noise is negligible.
Assume that uplink channel and downlink channel follow the channel reciprocity \cite{guillaud2005practical}.
The total energy harvested by MTCD $j$ served by MTCG $i$ can be evaluated as
\begin{equation}\label{sys1eq5}
E_{ij}^{\text H}=\sum_{k=1}^K t_{N+k}u\left(\sum_{n=I_{k-1}+1}^{ I_k}|h_{nj}|^2 q_n \right), \quad \forall i\in\mathcal N, j \in \mathcal J_i,
\end{equation}
where $\sum_{n=I_{k-1}+1}^{ I_k}|h_{nj}|^2 q_n$ is the received RF power of MTCD $j$ during time $t_{N+k}$, and
function $u(\cdot)$ captures the EH model which maps input RF power into harvested power.
Two commonly used EH models are shown in Fig.~\ref{syseh}, i.e., linear and non-linear EH models.
According to \cite{7264986} and \cite{7511403}, linear EH model  may lead to resource allocation mismatch.
In order to capture the effects of practical EH circuits on the end-to-end power conversion, we adopt the more practical non-linear EH model proposed in \cite{7264986}:
\begin{equation}\label{sys1eq5_2}
u(x)=\left\{ \begin{array}{ll}
\frac{M(1+\text e^{ab})}{\text{e} ^{ab}+\text{e} ^{-a(x-2b)}}
-\frac{M}{\text{e}^{ab}}, & \text{if $x\geq P_0$}
\\
0, & \text{elsewise}
\end{array}
\right. ,
\end{equation}
where $a$, $b$, $M$ and $P_0$ are positive parameters which capture the joint effects of different non-linear phenomena caused by hardware constraints.
Note that $P_0$ is the receiver sensitivity threshold of each MTCD, in which energy conversion starts.
Hence, it is possible that some MTCDs cannot effectively harvest energy in some slots, since the received power is below the receiver sensitivity threshold $P_0$.

The total energy consumption of the M2M enabled communication network consists of two parts: the energy consumed by MTCDs and MTCGs.
For each part, the energy consumption of a transmitter consists of both RF transmission power and circuit power due to hardware processing \cite{Wu2015Resource}.
According to \cite{Wu2016Energy},
the energy consumption when MTCDs or MTCGs are in idle model, i.e., do not transmit RF signals, is negligible.

During the $i$-th ($i\leq N$) phase, MTCD $j \in \mathcal J_i$ served by MTCG $i$ just transmits data to MTCG $i$ with allocated time $t_i$ and transmission power
$p_{j}$.
Thus, the energy $E_{ij}$ consumed by MTCD $j \in \mathcal J_i$ can be modeled as
\begin{equation}\label{sys1eq6}
E_{ij}= t_i \left(\frac{p_j}{\eta} + P^{\text C}\right),
\quad \forall i\in\mathcal N, j \in \mathcal J_i,
\end{equation}
%
%the system energy consumption, denoted by $E_1$, is modeled as
%\begin{equation}\label{sys1eq6}
%E_1=\sum_{i \in \mathcal N}^{j \in \mathcal J_i} t_i \left(\frac{p_j}{\eta} + P^{\text C}\right),
%\end{equation}
where $\eta\in(0,1]$ and $P^{\text C}$ denote the power amplifier (PA) efficiency and the circuit power consumption of each MTCD, respectively.
According to the energy causality constraint in EH networks,
$E_{ij}$ has to satisfy $E_{ij} \leq E_{ij}^{\text H}$.
Summing the energy consumed by all MTCDs in $\mathcal J_i$, we can obtain the energy $E_i$ consumed during the $i$-th phase as
\begin{equation}\label{sys1eq6_2}
E_i=\sum_{j=J_{i-1}+1}^{J_i}E_{ij},
\quad \forall i\in\mathcal N.
\end{equation}

During the ($N+k$)-th phase, the system energy consumption, denoted by $E_{N+k}$, is modeled as
\begin{eqnarray}\label{sys1eq7}
E_{N+k}&&\!\!\!\!\!\!\!\!\!\!=\!\sum_{i =I_{k-1}+1}^{ I_k} t_{N+k} \left(\frac{q_i}{\xi} + Q^{\text C}\right)
\nonumber\\
&&\!\!\!\!\!\!\!\!\!\!
\quad-
\sum_{i=1}^{N} \sum_{j=J_{i-1}+1}^{ J_i} t_{N+k}u\left(\sum_{n=I_{k-1}+1}^{ I_k}|h_{nj}|^2 q_n \right)
%\nonumber\\
%&&\!\!\!\!\!\!\!\!\!\!=\sum_{i =1}^{ N} t_{N+1} \left(\frac{q_i}{\xi} + Q^{\text C}\right)
%-\sum_{i=1}^{ N} \sum_{j=J_{i-1}+1}^{ J_i}\sum_{n=1}^{ N} \zeta t_{N+1}q_n |h_{nj}|^2
,
\end{eqnarray}
where $\xi\in(0,1]$ and $Q^{\text C}$ are the PA efficiency and the circuit power consumption of each MTCG, respectively.
According to the law of energy conservation \cite{Ng2013Wireless},
we must have
\begin{equation}\label{sys1eq7_2}
\sum_{i=I_{k-1}+1}^{I_k}\!\! t_{N+k} q_i \!-\!\sum_{i=1}^{ N} \!\sum _{j=J_{i-1}+1}^{J_i} t_{N+k}u\!\left(\!\sum_{n=I_{k-1}+1}^{ I_k}\!|h_{nj}|^2 q_n \!\right)\!\!>\!0,
\end{equation}
which is the energy loss due to wireless propagation.

Based on (\ref{sys1eq5})-(\ref{sys1eq7}), the total energy consumption, $E_{\text {Tot}}$, of the whole system during time $T$ can be expressed as
\begin{eqnarray}\label{sys1eq8}
\!\!\!\!\!\!E_{\text {Tot}}
&&\!\!\!\!\!\!\!\!\!\!=\sum_{i=1}^{N+K} E_i
\nonumber\\
&&\!\!\!\!\!\!\!\!\!\!=\sum_{i=1}^{ N}\sum_{j=J_{i-1}+1}^{ J_i}t_i \left(\frac{p_j}{\eta} + P^{\text C}\right)
\nonumber\\
&&\!\!\!\!\!\!\!\!\!\!
\quad+
\sum_{k=1}^{K}\sum_{i=I_{k-1}+1}^{I_k} t_{N+k} \left(\frac{q_i}{\xi} + Q^{\text C}\right)
\nonumber\\
&&\!\!\!\!\!\!\!\!\!\!
\quad-
\sum_{k=1}^K
\sum_{i=1}^{ N} \sum_{j=J_{i-1}+1}^{ J_i}t_{N+k}u\left(\!\sum_{n=I_{k-1}+1}^{I_k} |h_{nj}|^2 q_n\!\right)\!\!.
\end{eqnarray}

\subsection{TDMA Strategy}

\begin{figure}
\centering
\includegraphics[width=3.35in]{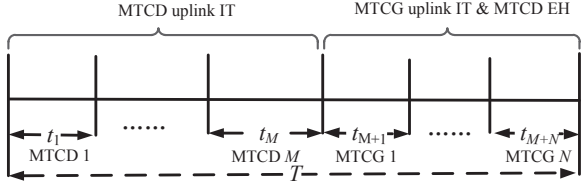}
%\vspace{-1.8em}
\caption{Time sharing scheme for TDMA strategy during one uplink transmission period.}\label{sys5}
%\vspace{-1.8em}
\end{figure}

With the TDMA strategy, time $T$ consists of $M+N$ uplink transmission phases for MTCDs and MTCGs, as illustrated in Fig.~\ref{sys5}.
%TDMA strategy is adopted for each MTCD and MTCG to upload the transmission data.
All MTCDs transmit data to the corresponding MTCGs in the first $M$ phases with TDMA, and all MTCGs transmit the collected data to the BS in the last $N$ phases with TDMA.
Then, we obtain the following transmission time constraint
\begin{equation}\label{sys1eq1tdma}
\sum_{i =1} ^{M+N}t_i \leq T.
\end{equation}

In the $j$-th ($j\leq M$) phase, MTCD $j\in \mathcal J_i$ transmits data to its serving MTCG $i$ with achievable throughput
\begin{equation}\label{sys1eq3tdma}
r_{ij}=B t_{j}\log_2\left(
1+\frac{|h_{ij}|^2 p_j}
{ \sigma^2}
\right), \quad \forall i \in \mathcal N, j \in \mathcal J_i.
\end{equation}
In the $(M+i)$-th phase, after having decoded all the messages of its served MTCDs, MTCG $i$ transmits the collected data to the BS with achievable throughput
\begin{equation}\label{sys1eq4tdma}
r_i=B t_{M+i}\log_2\left(
1+\frac{|h_{i}|^2 q_i}
{\sigma^2}
\right),\quad \forall i \in \mathcal N.
\end{equation}

Similar to (\ref{sys1eq5}), the total energy harvested of MTCD $j$ served by MTCG $i$ is
\begin{equation}\label{sys1eq5tdma}
E_{ij}^{\text H}= \sum_{n=1}^{ N}  t_{M+n}u(|h_{nj}|^2 q_n ), \quad \forall i\in\mathcal N, j \in \mathcal J_i.
\end{equation}
According to (\ref{sys1eq5_2}), it is possible that some MTCDs cannot effectively harvest energy in some slots, due to the fact that the received power is below the receiver sensitivity threshold $P_0$.

As in (\ref{sys1eq6}) and (\ref{sys1eq7}),
the energy consumption of a transmitter includes both RF transmission power and circuit power \cite{Wu2015Resource}. % \cite{Ng2013Wireless,Miao2013Energy,Wu2015Resource,Huang2013Simultaneous}.
With allocated transmission time $t_j$, the energy $E_{ij}$ consumed by MTCD $j \in \mathcal J_i$ can be modeled as
\begin{equation}\label{sys1eq6tdma}
E_{ij}= t_j \left(\frac{p_j}{\eta} + P^{\text C}\right),
\quad \forall i\in\mathcal N, j \in \mathcal J_i.
\end{equation}
With allocated transmission time $t_{M+i}$,
the system energy consumption, denoted by $E_{M+i}$, is modeled as
\begin{equation}\label{sys1eq7tdma}
E_{M+i}= t_{M+i} \left(\frac{q_i}{\xi} + Q^{\text C}\right)-\sum_{n=1}^{N} \sum_{j=J_{n-1}+1}^{ J_n} t_{M+i} u(|h_{ij}|^2 q_i ),
\end{equation}
where $ \sum_{n=1}^{N} \sum_{j=J_{n-1}+1}^{ J_n} t_{M+i}u(q_i |h_{ij}|^2)$ is the energy harvested by all MTCDs during the transmission time $t_{M+i}$ for MTCG $i$ to transmit data to the BS.

According to (\ref{sys1eq5tdma})-(\ref{sys1eq7tdma}), the total energy consumption, $E_{\text {Tot}}$, of the whole system during time $T$ can be expressed as
\begin{eqnarray}\label{sys1eq8tdma}
E_{\text {Tot}}
&&\!\!\!\!\!\!\!\!\!\!=\sum_{i=1}^{ N}\sum_{j=J_{i-1}+1}^{ J_i} E_{ij}+
\sum_{i=M+1}^{M+N} E_i
\nonumber\\
&&\!\!\!\!\!\!\!\!\!\!=\sum_{i=1}^{ N}\sum_{j=J_{i-1}+1}^{ J_i}t_j \left(\frac{p_j}{\eta} + P^{\text C}\right)+\sum_{i=1}^{N} t_{M+i} \left(\frac{q_i}{\xi} + Q^{\text C}\right)
\nonumber\\
&&\!\!\!\!\!\!\!\!\!\!
\quad-\sum_{i=1}^{ N} \sum_{j=J_{i-1}+1}^{ J_i}\sum_{n=1}^{ N}  t_{M+n}u( |h_{nj}|^2 q_n).
\end{eqnarray}

%% 20170921
\section{Energy Efficient Resource Allocation for NOMA}

In this section, we study the resource allocation for an uplink M2M enabled cellular network with NOMA and EH.
Specifically, we aim at minimizing the total energy consumption via jointly optimizing power control and time allocation for NOMA.
The system energy minimization problem is formulated as
\begin{subequations}\label{ee2min1}
\begin{align}
\!\!\mathop{\min}_{\pmb{p}, \pmb q, \pmb t}\:\;
&  E_{\text{Tot}}
%\sum_{i=1}^{ N}\sum_{j=J_{i-1}+1}^{ J_i}t_i \left(\frac{p_j}{\eta} + P^{\text C}\right)+\sum_{i=1}^{N} t_{N+1} \left(\frac{q_i}{\xi} + Q^{\text C}\right)
%\nonumber\\
%&-\sum_{i=1}^{ N} \sum_{j=J_{i-1}+1}^{ J_i}\sum_{n=1}^{ N} \zeta t_{N+1}q_n |h_{nj}|^2
\\
\textrm{s.t.}\qquad \!\!\!\!\!\!\!\!
&r_{ij} \geq D_j, \quad\forall i\in \mathcal N, j \in \mathcal J_i\\
&r_i  \geq  \sum_{j=J_{i-1}+1}^{ J_i} D_j, \quad\forall i \in \mathcal N \\
&E_{ij} \leq E_{ij}^{\text H}, \quad\forall i\in \mathcal N, j \in \mathcal J_i\\
&\sum_{i=1}^{N+K} t_i \leq T \\
&0\leq p_j \leq P_j, 0\leq q_i \leq Q_i, \quad\forall i\in \mathcal N, j \in \mathcal J_i\\
&\pmb t\geq \pmb 0,
\end{align}
\end{subequations}
where
$\pmb p=[p_1, \cdots, p_M]^T$,
$\pmb q=[q_1, \cdots, q_N]^T$,
$\pmb t=[t_{1}, \cdots,$ $ t_{N+K}]^T$,
$D_j$ is the payload that MTCD $j$ has to upload within time constraint $T$,
$P_{j}$ is the maximal transmission power of MTCD $j$,
and $Q_{i}$ is the maximal transmission power of MTCG $i$.
It is assumed that all payloads are positive, i.e., $D_j>0$, for all $j$.
The objective function (\ref{ee2min1}a) defined in (\ref{sys1eq8}) is the total energy consumption of both MTCDs and MTCGs.
Constraints (\ref{ee2min1}b) and (\ref{ee2min1}c) reflect that the minimal required payloads for MTCDs can be uploaded to the BS.
The consumed energy of each MTCD should not exceed its harvested energy in time $T$, as stated in (\ref{ee2min1}d).
Constraints (\ref{ee2min1}e) reflect that the payloads for all MTCDs are transmitted in time $T$.

Note that problem (\ref{ee2min1}) is nonconvex due to nonconvex objective function (\ref{ee2min1}a) and constraints (\ref{ee2min1}b)-(\ref{ee2min1}d).
In general, there is no standard algorithm for solving nonconvex optimization problems.
In the following, we first find the optimal conditions for problem (\ref{ee2min1}) by exploiting the special structure of the uplink NOMA rate, and then provide an iterative power control and time allocation algorithm.

\subsection{Optimal Conditions}
By analyzing problem (\ref{ee2min1}), we have the following lemma.
\begin{lemma}
The optimal solution ($\pmb p^*, \pmb q^*, \pmb t^*$) to problem (\ref{ee2min1}) satisfies
\begin{equation}\label{ee2eq1}
r_{ij}^*=D_j, \quad \forall i \in \mathcal N, j \in \mathcal J_i.
\end{equation}
\end{lemma}

This observation states that the minimal throughput leads to more energy saving, which is similar to \cite{Yang2017Energy} and is also widely known in the information theory community.
%This observation is reasonable as less resource is utilized and hence less energy is consumed.
%Intuitively, we can verify that $r_{ij}^*=D_j$, as otherwise objective function (\ref{ee2min1}a) can be further improved with all constraints satisfied by decreasing transmission power $p_j^*$, contradicting that the solution is optimal.

Lemma 1 states that the optimal transmit throughput for each MTCD is required minimum.
Note that the optimal throughput for each MTCG is not always as its minimum requirement, i.e., constraints (\ref{ee2min1}c) are active at the optimum, since MTCGs should transmit more power to maintain that the harvested energy of each MTCD is no less than the consumed energy.

Based on Lemma 1, we further have the following lemma about the optimal transmission power of MTCDs.
\begin{lemma}
If ($\pmb p^*, \pmb q^*, \pmb t^*$) is the optimal solution to problem (\ref{ee2min1}), we have
\begin{eqnarray}\label{ee2eq2}
 p_{j}^* &&\!\!\!\!\!\!\!\!\!\!=
 \sum_{l=j+1}^{J_i}
\frac{\sigma^2}
{|h_{ij}|^2 }
 {\left( {\text e}^{\frac{a_{l}}{t_i^*}} - 1 \right)
 \left( {\text e^{\frac{a_{j}}{t_i^*}}} -  1 \right)
\text e^{ \frac{b_{jl}}{  t_i^*}}}
\nonumber\\
&&\!\!\!\!\!\!\!\!\!\!\quad+
\frac{\sigma^2}
{|h_{ij}|^2 }
{\left(\text e^{\frac{a_{j}}{t_i^*}} -1\right)
}
,\quad \forall i \in \mathcal N, j \in \mathcal J_i,
\end{eqnarray}
where
\begin{equation}\label{ee2eq2_2}
a_l={\frac{(\ln 2)D_{l}}{B}},
\quad b_{jl}={\frac{\sum_{s=j+1}^{l-1} (\ln 2) D_{s}}{B}},
\quad \forall i\in \mathcal N, j,l \in \mathcal J_i.
\end{equation}
Besides, the optimal transmission power $p_j^*$ of MTCD $j\in \mathcal J_i$ is always non-negative and decreases with the transmission time $t_i^*$.
\end{lemma}

\itshape \textbf{Proof:}  \upshape Please refer to Appendix A.
\hfill $\Box$

From Lemma 2, large transmission time results in low transmission power.
This is reasonable as the minimal payload is limited and large transmission time requires low achievable rate measured in bits/s.
It is also revealed from Lemma 2 that the optimal transmission power of MTCD $j \in \mathcal J_i$ served by MTCG $i$ depends only on the variable of the allocated transmission time $t_i$.
As a result, the energy $E_{ij}$ consumed by MTCD $j$ in $\mathcal J_i$ is a function of the allocated transmission time $t_i$.
Based on (\ref{sys1eq6}) and (\ref{ee2eq2}), we have
\begin{eqnarray}\label{ee2eq3}
E_{ij}&&\!\!\!\!\!\!\!\!\!\!=
\sum_{l=j+1}^{J_i}
\frac{\sigma^2 t_i}
{\eta |h_{ij}|^2 }
{\left( {\text e}^{\frac{a_{l}}{t_i}} - 1 \right)
 \left( {\text e^{\frac{a_{j}}{t_i}}} -  1 \right)
\text e^{ \frac{b_{jl}}{  t_i}}}
 \nonumber\\
 &&\!\!\!\!\!\!\!\!\!\! \quad+
\frac{\sigma^2 t_i}
{\eta |h_{ij}|^2 }
{\left(\text e^{\frac{a_{j}}{t_i}} -1\right)
}
%\nonumber\\
%&&\!\!\!\!\!\!\!\!\!\! \quad
+t_i P_{\text C}.
\end{eqnarray}

\begin{theorem}
The energy $E_{ij}$ defined in (\ref{ee2eq3}) is convex with respect to (w.r.t.) the transmission time $t_i$.
When $P^{\text C}=0$, the energy $E_{ij}$ monotonically decreases with $t_i$.
When $P^{\text C}>0$,
the energy $E_{ij}$ first decreases with $t_i$ when $0\leq t_i \leq T_{ij}^*$ and then increases with $t_i$ when $t_i>T_{ij}^*$, where $T_{ij}^*$ is the unique zero point of the first-order derivative $\frac {\partial E_{ij}}{\partial t_i}$, i.e.,
\begin{eqnarray}\label{ee2eq5}
\left.\frac {\partial E_{ij}}{\partial t_i}
\right|_{t_i=T_{ij}^*}
=0.
\end{eqnarray}
\end{theorem}

\itshape \textbf{Proof:}  \upshape Please refer to Appendix B.
\hfill $\Box$
\begin{figure}
\centering
\includegraphics[width=3in]{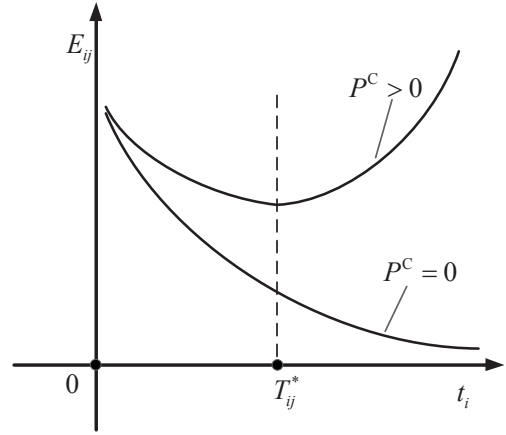}
%\vspace{-1em}
\caption{The energy $E_{ij}$ versus transmission time $t_i$.}\label{sys3}
%\vspace{-1em}
\end{figure}

Fig.~\ref{sys3} exemplifies the energy $E_{ij}$ given in (\ref{ee2eq3}) versus $t_i$.
When $P^{\text C}=0$, i.e., the circuit energy consumption of MTCDs is not considered,
we come to the same conclusion as in \cite{angelakis2014minimum} and \cite{Chin2015Power} that the consumed energy decreases as the transmission time increases according to Theorem 1.
Without considering the circuit power consumption, $R=\log_2(1+\text{SNR})$ and the energy efficiency increases with the decrease of power.
Consequently, when $P^{\text C}=0$ and the minimal throughput demand $D_j$ is given, the consumed energy $E_{ij}$ is a decreasing function w.r.t. $t_i$.
This fundamentally follows the Shannon's law.
%energy saving can always benefit from longer transmission time.

When $P^{\text C}>0$, i.e., the circuit energy consumption of MTCDs is taken into account, however, we find from Theorem 1 that the consumed energy first decreases and then increases with the transmission time, which is different from the previous conclusion in \cite{angelakis2014minimum} and \cite{Chin2015Power}.
This is because that the total energy contains two parts balancing each other, i.e., the RF transmission energy part which monotonically decreases with the transmission time and the circuit energy part which linearly increases with the transmission time.
In the following of this section, we assume that the circuit power consumption of MTCDs and MTCGs is in general positive, i.e., $P^{\text C}>0$ and $Q^{\text C}>0$.

\begin{theorem}
If $T\leq \max_{\forall i \in \mathcal N}\min_{\forall j \in \mathcal J_i} \{T_{ij}^*\}$, the optimal time allocation $\pmb t^*$ to problem (\ref{ee2min1}) satisfies
\begin{equation}\label{ee2eq6}
\sum_{i=1}^{N+K}t_i^*=T.
\end{equation}
If $T \geq T_{\text{Upp}}$, where $T_{\text{Upp}}$ is defined in (\ref{appenCeqn6}), the optimal time allocation $\pmb t^*$ to problem (\ref{ee2min1}) satisfies
%\begin{equation}\label{ee2eq6_2}
%t_i^*=T_i^*, \quad \forall i \in \mathcal N,
%\end{equation}
%and
\begin{equation}\label{ee2eq6_3}
\sum_{i=1}^{N+K}t_i^*<T.
\end{equation}
\end{theorem}

\itshape \textbf{Proof:}  \upshape Please refer to Appendix C.
\hfill $\Box$

From Theorem 2, it is observed that transmitting with the maximal transmission time $T$ is optimal when $T$ is not large.
This is because that the reduced energy of RF transmission dominates the additional energy of circuit by increasing transmission time.
When the available time $T$ becomes large enough,
Theorem~2 states that
%the optimal transmission time for each group of MTCDs is unique, i.e., the optimal transmission time $t_i$ ($i \leq N$) does not increase with the maximal transmission time $T$.
%This is due to the fact the energy $E_i$ consumed by all MTCDs in $\mathcal J_i$ is a convex function which first decreases and then increases with $t_i$ from Theorem~1.
%According to Theorem 2,
it is not optimal to transmit with the maximal transmission time $T$.
This is due to the fact that the increased energy of circuit power dominates the power consumption while the energy reduction of RF transmission becomes relatively marginal.

\subsection{Joint Power Control and Time Allocation Algorithm}
Problem (\ref{ee2min1}) has two difficulties: one comes from the non-smooth EH function defined in (\ref{sys1eq5_2}),
and the other one is the non-convexity of both  objective function (\ref{ee2min1}a) and constraints (\ref{ee2min1}b)-(\ref{ee2min1}d).
To deal with the first difficulty, we introduce notation $\mathcal S_{ij}$ as the set of phases during which MTCD $j\in\mathcal J_i$ can effectively harvest energy, i.e., $\mathcal S_{ij}=\{k|\sum_{n=I_{k-1}+1}^{I_k}|h_{nj}|^2q_n>P_0, \forall k \in \mathcal K\}$.
With $\mathcal S_{ij}$ in hand, the harvested power of MTCD $j\in\mathcal J_{i}$ can be presented by the smooth function $\bar u(x)$ defined in (\ref{sys1eq5_22}).
To deal with the second difficulty,
we substitute (\ref{sys1eq3})-(\ref{sys1eq6}), (\ref{sys1eq8}) and (\ref{ee2eq2}) into (\ref{ee2min1}), and the original problem (\ref{ee2min1})  with fixed sets $\mathcal S_{ij}$'s can be equivalently transformed into the following problem:
\begin{subequations}\label{ee2min2}
\begin{align}
\!\!\mathop{\min}_{\pmb q, \pmb t}\;
\!&  \sum_{i=1}^{ N}\sum_{j=J_{i-1}+1}^{ J_i}
\sum_{l=j+1}^{J_i}
\frac{\sigma^2 t_i}
{\eta |h_{ij}|^2 }
{\left( {\text e}^{\frac{a_{l}}{t_i}} - 1 \right)
 \left( {\text e^{\frac{a_{j}}{t_i}}} -  1 \right)
\text e^{ \frac{b_{jl}}{  t_i}}}
\nonumber \\
 & +\!\sum_{i=1}^{ N}\!\sum_{j=J_{i-1}+1}^{ J_i}\!
\frac{\sigma^2 t_i}
{\eta |h_{ij}|^2 }
{\!\left(\!\text e^{\frac{a_{j}}{t_i}} \!-\!1\!\right)\!
}
%\nonumber\\
%&
\!+\!\sum_{i=1}^{ N}\!\sum_{j=J_{i-1}+1}^{ J_i}\!t_i P^{\text C}
\nonumber\\
&+\sum_{k=1}^K\sum_{i=I_{k-1}+1}^{I_k} t_{N+k} \left(\frac{q_i}{\xi} + Q^{\text C}\right)
\nonumber\\
&-\sum_{i=1}^{ N} \sum_{j=J_{i-1}+1}^{ J_i}\sum_{k\in\mathcal S_{ij}}t_{N+k} \bar u \left(\sum_{n=I_{k-1}+1}^{I_k} |h_{nj}|^2 q_n\right)
\\
\textrm{s.t.}\qquad \!\!\!\!\!\!\!\!\!
%&B t_{N+1}\log_2\!\!\left(\!\!
%1\!+\!\frac{|h_{i}|^2 q_i}
%{ \sum_{n=i+1}^{N}\! \!|h_{n}|^2 q_n\!+\!\sigma^2}
%\!\!\right)\!
%%\nonumber\\
%%&
%\!\geq \! \sum_{j\in\mathcal J_i} \!D_j,  \!\forall i \!\in\! \mathcal N \\
&{|h_{i}|^2 q_i}\geq \!\left(\!2^{\frac{\sum_{j\in\mathcal J_i} D_j}{B t_{N+1}}}\!-\!1\!\right)
\!\left({ \sum_{l=i+1}^{I_k} |h_{l}|^2 q_l\!+\!\sigma^2}\!\right),
\nonumber\\
&\qquad\qquad\forall k \in \mathcal K, i \in \mathcal I_k\\
& \sum_{l=j+1}^{J_i}
\frac{\sigma^2}
{\eta |h_{ij}|^2 }
 {t_i \left( {\text e}^{\frac{a_{l}}{t_i}} - 1 \right)
 \left( {\text e^{\frac{a_{j}}{t_i}}} -  1 \right)
\text e^{ \frac{b_{jl}}{  t_i}}}
\nonumber\\
&+
\frac{\sigma^2}
{\eta |h_{ij}|^2 }
{t_i\left(\text e^{\frac{a_{j}}{t_i}} -1\right)
} +t_i P^{\text C}
 \nonumber\\
 &\leq \! \sum_{k\in\mathcal S_{ij}}\!t_{N+k} \bar u\!\left(\!\sum_{n=I_{k-1}+1}^{ I_k} \! |h_{nj}|^2 q_n\!\right),\! \quad\forall i\in \mathcal N, j \in \mathcal J_i\\
 &\sum_{n=I_{k-1}+1}^{ I_k}\! |h_{nj}|^2 q_n\!\geq\! P_0, \!\!\!\quad \forall i \in \mathcal N, j \in \mathcal J_i, k \in \mathcal S_{ij}\\
% &\sum_{n=1}^{ N} |h_{nj}|^2 q_n\geq P_0, \quad\forall i\in \mathcal N, j \in \mathcal J_i\\
&\sum_{i=1}^{N+K} t_i \leq T \\
&\sum_{l=j+1}^{J_i}
\frac{\sigma^2}
{|h_{ij}|^2 }
 {\left( {\text e}^{\frac{a_{l}}{t_i}} - 1 \right)
 \left( {\text e^{\frac{a_{j}}{t_i}}} -  1 \right)
\text e^{ \frac{b_{jl}}{  t_i}}}
\nonumber\\
&  +
\frac{\sigma^2}
{|h_{ij}|^2 }
{\left(\text e^{\frac{a_{j}}{t_i}} -1\right)
}
%\nonumber\\
%&
\leq P_{j}, \quad\forall i\in \mathcal N, j \in \mathcal J_i\\
& 0\leq q_i \leq Q_i, \quad\forall i\in \mathcal N\\
&\pmb t \geq \pmb 0,
\end{align}
\end{subequations}
where \begin{equation}\label{sys1eq5_22}
\bar u(x)=
\frac{M(1+\text e^{ab})}{\text{e} ^{ab}+\text{e} ^{-a(x-2b)}}
-\frac{M}{\text{e}^{ab}}, \quad \forall x\geq 0.
\end{equation}

%Note that constraints (\ref{ee2min2}d) follow from that input RF power $\sum_{n=1}^{ N} q_n |h_{nj}|^2$ should be more than the receiver sensitivity threshold $P_0$ to satisfy (\ref{ee2min2}c).
%Denote $\pmb t_{-i}=(t_1, \cdots, t_{i-1}, t_{i+1}, \cdots, t_{N+1})^T$.
Problem (\ref{ee2min2}) is still nonconvex w.r.t. ($\pmb q, \pmb t$) due to nonconvex objective function (\ref{ee2min2}a) and constraints (\ref{ee2min2}b)-(\ref{ee2min2}c).
Before solving problem (\ref{ee2min2}), we have the following theorem.

\begin{theorem}
Given transmission time $\pmb \tau=[t_{N+1}, \cdots,$ $t_{N+K}]^T$,  problem (\ref{ee2min2}) is a convex problem w.r.t. ($\pmb q, \bar {\pmb t}$), where $\bar{\pmb t}=[t_1, \cdots, t_{N}]^T$.
Given ($\pmb q, \bar {\pmb t}$), problem (\ref{ee2min2}) is equivalent to a linear problem w.r.t. $\pmb \tau$.
\end{theorem}

\itshape \textbf{Proof:}  \upshape Please refer to Appendix D.
\hfill $\Box$

According to Theorem 3, problem (\ref{ee2min2}) with given transmission time $\pmb \tau$ can be effectively solved by using the standard convex optimization method, such as interior point method \cite{boyd2004convex}.
Besides, problem (\ref{ee2min2}) with given  ($\pmb q, \bar {\pmb t}$) is a linear problem, which can be optimally solved via the simplex method.
%
%As a result, problem (\ref{ee2min2}) can be solved to its global optimality by one-dimensional exhaustive search for transmission time $t_{N+1}$.
Based on Theorem 3, we propose an iterative power control and time allocation for NOMA (IPCTA-NOMA) algorithm with low complexity to obtain a suboptimal solution of problem (\ref{ee2min1}).
The idea is to iteratively update sets $\mathcal S_{ij}$'s according to the power and time variables obtained in the previous step.

\begin{algorithm}[h]
\caption{\!\!: Iterative Power Control and Time Allocation for NOMA (IPCTA-NOMA) Algorithm}
\begin{algorithmic}[1]
\STATE Set $\mathcal S_{ij}^{(0)}=\{k|j\in \cup_{n\in\mathcal I_k}\mathcal J_n, k \in \mathcal K\}$, $\forall i \in \mathcal I$, $j \in \mathcal J_{ij}$,
initialize a feasible solution ($\pmb q^{(0)}, {\pmb t}^{(0)}$) to problem (\ref{ee2min2}) with $\mathcal S_{ij}^{(0)}$'s, the tolerance $\theta$, the iteration number $v=0$, and the maximal iteration number $V_{\max}$.
\REPEAT
\STATE Set $\pmb \tau^*=[t_{N+1}^{(v)}, \cdots, t_{N+K}^{(v)}]^T$.
\REPEAT
\STATE Obtain the optimal ($ {\pmb  q}^{*}, \bar {\pmb t}^{*}$) of convex problem  (\ref{ee2min2}) with fixed $\pmb \tau^*$ and sets $\mathcal S_{ij}^{(v)}$.
\STATE Obtain the optimal $\pmb \tau^*$ of linear problem  (\ref{ee2min2}) with fixed ($ {\pmb  q}^{*}, \bar {\pmb t}^{*}$) and sets $\mathcal S_{ij}^{(v)}$.
\UNTIL the objective value (\ref{ee2min2}a) with fixed sets $\mathcal S_{ij}^{(v)}$  converges.
\STATE Set $v=v+1$.
\STATE Denote $\pmb q^{(v)}=\pmb q^*$, $\pmb t^{(v)}=[\bar{\pmb t}^{*T}, \pmb \tau^{*T}]^T$.
%Obtain the suboptimal solution ($\hat {\pmb p}^{(v)}, \hat {\pmb  q}^{(v)}, \hat {\pmb t}^{(v)}$) of nonconvex problem  (\ref{ee2min2}) with fixed sets $\mathcal S_{ij}^{(v)}$.
\STATE Calculate the objective value (\ref{ee2min2}a) with fixed sets $\mathcal S_{ij}^{(v)}$ as $U_{\text {Obj}}^{(v)}=E_{\text {Tot}}({\pmb  q}^{(v)}, {\pmb t}^{(v)})$.
\STATE Update $\mathcal S_{ij}^{(v)}=\{k|\sum_{n=I_{k-1}+1}^{I_k}|h_{nj}|^2q_n^{(v)}>P_0, \forall k \in \mathcal K\}$, $\forall i \in \mathcal I, j \in \mathcal J_i$.
\UNTIL $v \geq 1$ and $|U_{\text {Obj}}^{(v)}-U_{\text {Obj}}^{(v-1)}|/U_{\text {Obj}}^{(v-1)}<\theta$ or $v > V_{\max}$.
%\STATE Output ${\pmb p}^{*}=\hat {\pmb p}^{(v)}, \hat {\pmb t}^{*}=\hat {\pmb t}^{(v)}, q_{i} =\hat q_i / t_{M+i}, \forall i \in \mathcal I$.
\end{algorithmic}
\end{algorithm}

%\begin{algorithm}[h]
%\caption{: Energy-efficient CSS Power Adaptation algorithm}
%\begin{algorithmic}[1]
%\STATE given: the maximum iteration number $L_{max}$, the iteration index $i=0,j=0$, $\Delta\tau$ is the step-size of sensing time, $M=\frac{(T-kT_r)}{\Delta\tau}$ , and the error tolerance $\delta_1>0$, $\delta_2>0$;
%\STATE Initialization: $ Q_{de}=\alpha $, $\lambda^{(0)}=\lambda_0, \mu^{(0)}=\mu_0, \nu^{(0)}=\nu_0, \xi^{(0)}=\xi_0, s>0$;
%\FOR {$k=1:K$}
%\FOR {$m=1:M$}
%\STATE $\tau=m\Delta\tau$
%\WHILE  {$|F(\lambda^{(i)})\geq\delta_2|$ and $i\leq L_{max}$}
%\STATE calculate $P_0^*$ and $P_1^*$ by (19) and (20), respectively;
%\STATE update $\mu $, $ \nu$, and $\xi$ using subgradient method as follows:
%\REPEAT
%\STATE  $\mu^{(j+1)} = \big[ \mu^{(j)}-s(P_{av}-{(T-\tau-kT_r)} $
% $   \mathbb{E} \{{{q}_{0}{P}_{0}+{q}_{1}{P}_{1}}\}) \big]^{+} $;
%\STATE  $\nu^{(j+1)}=[\nu^{(j)}-s(Q_{av}-{(T-\tau-kT_r)} $
%$  \mathbb{E} \{{ [ {{P}({H}_{1})}{(1-{Q}_{de})}P_0+ {{P}({H}_{1})}{{Q}_{de}}P_1 ] g_{sp}}\})]^{+}$;
%\STATE  $\xi^{(j+1)}=[\xi^{(j)}-s(R_{av}-R_{min})]^{+}$;
%\STATE  $j=j+1;$
%\UNTIL $|\mu^{(j)}(P_{av}-{(T-\tau-kT_r)}\mathbb{E}\{{q_0P_0+q_1P_1}\})|$ $\leq\delta_1 $,
%$ |\nu^{(j)}(Q_{av}-{(T-\tau-kT_r)} \mathbb{E} \{ [ {P(H_1)}$ ${(1-Q_{de})}P_0+ {P(H_1)}{Q_{de}}P_1 ] g_{sp}\})|\leq \delta_1 $,   and $ |\xi^{(j)}(R_{av}-R_{min})|\leq \delta_1 $
%\STATE $ \lambda^{(i+1)}= \frac{{R}_{av}(P_0^*, P_1^*)} {{E}_{av}(P_0^*, P_1^*)}$;
%\STATE $ i=i+1$;
%\ENDWHILE
%\ENDFOR
%\ENDFOR
%\STATE return $ P_0=P_0^*, P_1=P_1^*$, and $\lambda^*= \lambda^{(i)}$.
%\end{algorithmic}
%\end{algorithm}

\subsection{Convergence and Complexity Analysis}
\begin{theorem}
Assuming $V_{\max}\rightarrow\infty$, the sequence (${\pmb  q}, {\pmb t}$) generated by the IPCTA-NOMA algorithm converges.
\end{theorem}

\itshape \textbf{Proof:}  \upshape Please refer to Appendix E.
\hfill $\Box$

According to the IPCTA-NOMA algorithm, the major complexity lies in solving the convex problem (\ref{ee2min2}) with fixed $\pmb \tau$.
Considering that the dimension of the variables in problem (\ref{ee2min2}) with fixed $\pmb \tau$ is $2N$,
the complexity of solving problem (\ref{ee2min2}) with fixed $\pmb \tau$ by using the standard
interior point method is $\mathcal O(N^3)$ \cite[Pages 487, 569]{boyd2004convex}.
As a result, the total complexity of the proposed IPCTA-NOMA algorithm is $\mathcal O(L_{\text{NO}}L_{\text{IT}}N^3)$, where $L_{\text{NO}}$ denotes the number of outer iterations of the IPCTA-NOMA algorithm, and $L_{\text{IT}}$ denotes the number of inner iterations of the IPCTA-NOMA algorithm for iteratively solving nonconvex problem (\ref{ee2min2}) with fixed sets $\mathcal S_{ij}$'s.

\section{Energy Efficient Resource Allocation for TDMA}
In this section, we study the energy minimization for the M2M enabled cellular network with TDMA.
According to (\ref{sys1eq3tdma})-(\ref{sys1eq6tdma}) and (\ref{sys1eq8tdma}), the energy minimization problem can be formulated as
\begin{subequations}\label{ee2min1tdma}
\begin{align}
\!\!\mathop{\min}_{\pmb{p}, \pmb q, \hat{\pmb t}}\;
\!&  \sum_{i=1}^{ N}\sum_{j=J_{i-1}+1}^{ J_i}t_j \left(\frac{p_j}{\eta} + P^{\text C}\right)+\sum_{i=1}^{N} t_{M+i} \left(\frac{q_i}{\xi} + Q^{\text C}\right)
\nonumber\\
&-\sum_{i=1}^{ N} \sum_{j=J_{i-1}+1}^{ J_i}\sum_{n=1}^{ N}   t_{M+n}u(|h_{nj}|^2 q_n )
\\
\textrm{s.t.}\qquad \!\!\!\!\!\!\!\!\!
&B t_{j}\log_2\left(
1+\frac{|h_{ij}|^2 p_j}
{ \sigma^2}
\right) \geq D_j, \quad\forall i\in \mathcal N, j \in \mathcal J_i\\
&B t_{M+i}\log_2\left(
1+\frac{|h_{i}|^2 q_i}
{\sigma^2}
\right)  \geq  \sum_{j=J_{i-1}+1}^{ J_i} D_j, \quad\forall i \in \mathcal N \\
& t_j \left(\frac{p_j}{\eta} + P^{\text C}\right) \leq \sum_{n=1}^{ N}   t_{M+n}u( |h_{nj}|^2 q_n), \quad\forall i\in \mathcal N, j \in \mathcal J_i\\
&\sum_{i=1}^{N+1} t_i \leq T \\
&0\leq p_j \leq P_j, 0\leq q_i \leq Q_i, \quad\forall i\in \mathcal N, j \in \mathcal J_i\\
&\hat{\pmb t} \geq \pmb 0,
\end{align}
\end{subequations}
where $\hat{\pmb t}=[t_{1}, \cdots,$ $ t_{M+N}]^T$.

Obviously, problem (\ref{ee2min1tdma}) is nonconvex due to nonconvex objective function (\ref{ee2min1tdma}a) and constraints (\ref{ee2min1tdma}b)-(\ref{ee2min1tdma}d).
In the following, we first provide the optimal conditions for problem (\ref{ee2min1tdma}), and then we propose a low-complexity algorithm to solve problem (\ref{ee2min1tdma}).
%show that problem (\ref{ee2min1tdma}) can be transformed into an equivalent convex problem, of which the globally optimal solution can be obtained.

\subsection{Optimal Conditions}
Similar to Lemma 1, it is also optimal for each MTCD to transmit with the minimal throughput requirement.
Accordingly, the following lemma is directly obtained.
\begin{lemma}
The optimal solution ($\pmb p^*, \pmb q^*, \hat{\pmb t}^*$) to problem (\ref{ee2min1tdma}) satisfies
\begin{equation}\label{ee2eq1tdma}
B t_{j}^*\log_2\left(
1+\frac{|h_{ij}|^2 p_j^*}
{ \sigma^2}
\right) = D_j, \quad \forall i \in \mathcal N, j \in \mathcal J_i.
\end{equation}
\end{lemma}

According to (\ref{ee2eq1tdma}), the optimal transmission power of MTCD $j$ can be presented as
\begin{equation}\label{ee2eq2tdma}
p_j=\frac 1 {|h_{ij}|^2}\left(2^{\frac{D_j}{B t_j}}-1
\right), \quad \forall i \in \mathcal N, j \in \mathcal J_i.
\end{equation}
Substituting (\ref{ee2eq2tdma}) into (\ref{sys1eq6tdma}) yields
\begin{equation}\label{ee2eq3tdma}
E_{ij}= \frac { t_j}  {|h_{ij}|^2 \eta }\left(2^{\frac{D_j}{B t_j}}-1\right)
   +  t_j P^{\text C} ,
\quad \forall i\in\mathcal N, j \in \mathcal J_i.
\end{equation}

By analyzing (\ref{ee2eq3tdma}), we can obtain the following theorem.
\begin{theorem}
The energy $E_{ij}$ defined in (\ref{ee2eq3tdma}) is convex w.r.t. $t_j$.
When $P^{\text C}=0$, the energy $E_{ij}$ monotonically decreases with the transmission time $t_j$.
When $P^{\text C}>0$,
the energy $E_{ij}$ first decreases with $t_j$ when $0\leq t_j \leq T_{ij}^*$ and then increases with $t_j$ when $t_j>T_{ij}^*$, where $T_{ij}^*$ is the unique zero point of the first-order derivative $\frac {\partial E_{ij}}{\partial t_j}$, i.e.,
\begin{eqnarray}\label{ee2eq5tdma}
\left.\frac {\partial E_{ij}}{\partial t_j}
\right|_{t_j=T_{ij}^*}
=0.
\end{eqnarray}
\end{theorem}

Since Theorem 5 can be proved by checking the first-order derivative $\frac {\partial E_{ij}}{\partial t_j}$ as in Appendix B, the proof of Theorem 4 is omitted.
Similar to Theorem 2 for NOMA, we come to the similar conclusion for TDMA that transmitting with the maximal transmission time $T$ is optimal when $T$ is not large, while for large $T$ it is not optimal to transmit with the maximal transmission time $T$.

\subsection{Iterative Power Control and Time Allocation Algorithm}
Similar to problem (\ref{ee2min1}), problem (\ref{ee2min1tdma}) has two difficulties: one is the non-smooth EH function in (\ref{sys1eq5_2}),
and the other is the nonconvex  objective function (\ref{ee2min1tdma}a) and constraints (\ref{ee2min1tdma}b)-(\ref{ee2min1tdma}d).
To deal with the first difficulty, we introduce notation $\mathcal S_{ij}$ as the set of MTCGs from which MTCD $j\in\mathcal J_i$ can effectively harvest energy, i.e., $\mathcal S_{ij}=\{n||h_{nj}|^2q_n>P_0, \forall n \in \mathcal I\}$.
With $\mathcal S_{ij}$ in hand, the harvested power of MTCD $j\in\mathcal J_{i}$ from MTCG $n\in\mathcal S_{ij}$ can be presented by the smooth function $\bar u(x)$ defined in (\ref{sys1eq5_22}).
To tackle the second difficulty, we show that problem (\ref{ee2min1tdma}) with fixed sets $\mathcal S_{ij}$'s can be transformed into an equivalent convex problem.

\begin{theorem}
The original problem in (\ref{ee2min1tdma}) with fixed sets $\mathcal S_{ij}$'s can be equivalently transformed into the following convex problem as
\begin{subequations}\label{ee2min2tdma}
\begin{align}
\!\!\mathop{\min}_{\hat {\pmb p}, \hat{\pmb q}, \hat{\pmb t}}\;
\!&  \sum_{i=1}^{ N}\sum_{j=J_{i-1}+1}^{ J_i} \left(\frac{\hat p_j}{\eta} +t_j P^{\text C}\right)+\sum_{i=1}^{N} \left(\frac{\hat q_i}{\xi} + t_{M+i}Q^{\text C}\right)
\nonumber\\
&-\sum_{i=1}^{ N} \sum_{j=J_{i-1}+1}^{ J_i}\sum_{n\in\mathcal S_{ij}} t_{M+n} \bar u\left(\frac{ |h_{nj}|^2\hat q_n }{t_{M+n}} \right)
\\
\textrm{s.t.}\qquad \!\!\!\!\!\!\!\!\!
&B t_{j}\log_2\left(
1+\frac{|h_{ij}|^2 \hat p_j}
{ \sigma^2 t_j}
\right) \geq D_j, \quad\forall i\in \mathcal N, j \in \mathcal J_i\\
&B t_{M+i}\log_2\left(
1+\frac{|h_{i}|^2 \hat q_i}
{\sigma^2 t_{M+i}}
\right)  \geq  \sum_{j=J_{i-1}+1}^{ J_i} D_j, \quad\forall i \in \mathcal N \\
& \frac{\hat p_j}{\eta} \!+\! t_j P^{\text C} \!\leq\!\! \sum_{n\in\mathcal S_{ij}} \! t_{M+n} \bar u\left(\!\frac{|h_{nj}|^2 \hat q_n }{t_{M+n}}\!\right), \!\!\quad\forall i\!\in\! \mathcal N, j \!\in\! \mathcal J_i\\
& {|h_{nj}|^2 \hat q_n } > P_0 {t_{M+n}}, \quad\forall i\in \mathcal N, j \in \mathcal J_i, n\in\mathcal S_{ij}\\
&\sum_{i=1}^{N+1} t_i \leq T \\
&0\leq \hat p_j \leq P_j t_j, \quad\forall i\in \mathcal N, j \in \mathcal J_i \\
&0\leq \hat q_i \leq Q_i t_{M+i}, \quad\forall i\in \mathcal N\\
&\hat {\pmb t} \geq \pmb 0,
\end{align}
\end{subequations}
where $\hat{\pmb p}=[\hat p_1, \cdots, \hat p_M]^T$ and
$\hat{\pmb q}=[\hat q_1, \cdots, \hat q_N]^T$.
\end{theorem}

\itshape \textbf{Proof:}  \upshape Please refer to Appendix F.
\hfill $\Box$

%% 1343
Based on Theorem 6, we propose an iterative power control and time allocation for TDMA (IPCTA-TDMA) algorithm with low complexity to obtain a suboptimal solution of problem (\ref{ee2min1tdma}).
The idea is to iteratively update sets $\mathcal S_{ij}$'s according to the power and time variables obtained in the previous step.

\begin{algorithm}[h]
\caption{\!\!: Iterative Power Control and Time Allocation for TDMA (IPCTA-TDMA) Algorithm}
\begin{algorithmic}[1]
\STATE Set $\mathcal S_{ij}^{(0)}=\{i\}$, $\forall i \in \mathcal I$, $j \in \mathcal J_{ij}$, the tolerance $\theta$, the iteration number $v=0$, and the maximal iteration number $V_{\max}$.
\REPEAT
\STATE Obtain the optimal ($\hat {\pmb p}^{(v)}, \hat {\pmb  q}^{(v)}, \hat {\pmb t}^{(v)}$) of convex problem  (\ref{ee2min2tdma}) with fixed sets $\mathcal S_{ij}^{(v)}$.
\STATE Calculate the objective value (\ref{ee2min2tdma}a) with fixed sets $\mathcal S_{ij}^{(v)}$ as $U_{\text {Obj}}^{(v)}=E_{\text {Tot}}(\hat {\pmb p}^{(v)}, \hat {\pmb  q}^{(v)}, \hat {\pmb t}^{(v)})$.
\STATE Set $v=v+1$.
\STATE Update $\mathcal S_{ij}^{(v)}=\left\{n\left|\frac{|h_{nj}|^2 \hat q_n^{(v-1)}}{t_{M+n}^{(v-1)}}>P_0\right., \forall n \in \mathcal I\right\}$, $\forall i \in \mathcal I, j \in \mathcal J_i$.
\UNTIL $v \geq 2$ and $|U_{\text {Obj}}^{(v)}-U_{\text {Obj}}^{(v-1)}|/U_{\text {Obj}}^{(v-1)}<\theta$ or $v > V_{\max}$.
\STATE Output ${\pmb p}^{*}=\hat {\pmb p}^{(v)}, \hat {\pmb t}^{*}=\hat {\pmb t}^{(v)}, q_{i}^* =\hat q_i^{(v)}/ t_{M+i}^{(v)}, \forall i \in \mathcal I$.
\end{algorithmic}
\end{algorithm}

%\begin{algorithm}[h]
%\caption{: Energy-efficient CSS Power Adaptation algorithm}
%\begin{algorithmic}[1]
%\STATE given: the maximum iteration number $L_{max}$, the iteration index $i=0,j=0$, $\Delta\tau$ is the step-size of sensing time, $M=\frac{(T-kT_r)}{\Delta\tau}$ , and the error tolerance $\delta_1>0$, $\delta_2>0$;
%\STATE Initialization: $ Q_{de}=\alpha $, $\lambda^{(0)}=\lambda_0, \mu^{(0)}=\mu_0, \nu^{(0)}=\nu_0, \xi^{(0)}=\xi_0, s>0$;
%\FOR {$k=1:K$}
%\FOR {$m=1:M$}
%\STATE $\tau=m\Delta\tau$
%\WHILE  {$|F(\lambda^{(i)})\geq\delta_2|$ and $i\leq L_{max}$}
%\STATE calculate $P_0^*$ and $P_1^*$ by (19) and (20), respectively;
%\STATE update $\mu $, $ \nu$, and $\xi$ using subgradient method as follows:
%\REPEAT
%\STATE  $\mu^{(j+1)} = \big[ \mu^{(j)}-s(P_{av}-{(T-\tau-kT_r)} $
% $   \mathbb{E} \{{{q}_{0}{P}_{0}+{q}_{1}{P}_{1}}\}) \big]^{+} $;
%\STATE  $\nu^{(j+1)}=[\nu^{(j)}-s(Q_{av}-{(T-\tau-kT_r)} $
%$  \mathbb{E} \{{ [ {{P}({H}_{1})}{(1-{Q}_{de})}P_0+ {{P}({H}_{1})}{{Q}_{de}}P_1 ] g_{sp}}\})]^{+}$;
%\STATE  $\xi^{(j+1)}=[\xi^{(j)}-s(R_{av}-R_{min})]^{+}$;
%\STATE  $j=j+1;$
%\UNTIL $|\mu^{(j)}(P_{av}-{(T-\tau-kT_r)}\mathbb{E}\{{q_0P_0+q_1P_1}\})|$ $\leq\delta_1 $,
%$ |\nu^{(j)}(Q_{av}-{(T-\tau-kT_r)} \mathbb{E} \{ [ {P(H_1)}$ ${(1-Q_{de})}P_0+ {P(H_1)}{Q_{de}}P_1 ] g_{sp}\})|\leq \delta_1 $,   and $ |\xi^{(j)}(R_{av}-R_{min})|\leq \delta_1 $
%\STATE $ \lambda^{(i+1)}= \frac{{R}_{av}(P_0^*, P_1^*)} {{E}_{av}(P_0^*, P_1^*)}$;
%\STATE $ i=i+1$;
%\ENDWHILE
%\ENDFOR
%\ENDFOR
%\STATE return $ P_0=P_0^*, P_1=P_1^*$, and $\lambda^*= \lambda^{(i)}$.
%\end{algorithmic}
%\end{algorithm}

\subsection{Convergence and Complexity Analysis}

\begin{theorem}
Assuming $V_{\max}\rightarrow\infty$, the sequence ($\hat {\pmb p}, \hat {\pmb  q}, \hat {\pmb t}$) generated by the IPCTA algorithm converges.
\end{theorem}

Theorem 7 can be proved by using the same method as in Appendix E. The proof of Theorem 7 is thus omitted.

According to the IPCTA-TDMA algorithm, the major complexity lies in solving the convex problem (\ref{ee2min2tdma}).
Considering that the dimension of the variables in problem (\ref{ee2min2tdma}) is $2(M+N)$,
the complexity of solving problem (\ref{ee2min2tdma}) by using the standard
interior point method is $\mathcal O((M+N)^3)$ \cite[Pages 487, 569]{boyd2004convex}.
As a result, the total complexity of the proposed IPCTA-TDMA algorithm is $\mathcal O(L_{\text{TD}}(M+N)^3)$, where $L_{\text{TD}}$ denotes the number of iterations of the IPCTA-TDMA algorithm.

\section{Numerical Results}
In this section, we evaluate the proposed schemes through simulations.
There are 40 MTCDs uniformly distributed with a BS in the center.
We adopt the ``data-centric'' clustering technique in \cite{6649207} for cluster formation of MTCDs,
the number of MTCGs is set as 12,
and the maximal number of MTCDs associated to one MTCG is 4.
For NOMA, all MTCGs are classified into 6 clusters based on the strong-weak scheme \cite{Mohammad17uplinkNOMA,7273963}.

The path loss model is $128.1+37.6\log_{10} d$ ($d$ is in km)
and the standard deviation of shadow fading
is $4$ dB \cite{access2010further}.
The noise power $\sigma^2=-104$ dBm, and the bandwidth of the system is $B=18$ KHz.
%The conversion efficiency of the EH process is set to $\zeta=0.9$,
%and the PA efficiencies of each MTCD and MTCG are set to $\eta=\xi=0.9$.
For the non-linear EH model in (\ref{sys1eq5_2}), we set $M=24$ mW, $a=1500$ and $b=0.0014$ according to \cite{7264986}, which are obtained by curve fitting from the measurement data in \cite{6259492}.
The receiver sensitivity threshold $P_0$ is set as $0.1$ mW.
The PA efficiencies of each MTCD and MTCG are set to $\eta=\xi=0.9$,
and the circuit power of each MTCG is $Q^{\text C}=500$ mW as in \cite{Wu2016Energy}.
We assume equal throughput demand for all MTCDs, i.e., $D_1=\cdots=D_M=D$, and equal maximal transmission power for each MTCD or MTCG, i.e., $P_1=\cdots=P_M=P$, and $Q_1=\cdots=Q_N=Q$.
Unless otherwise specified, parameters are set as $P=5$ mW, $P^{\text C}=0.5$ mW, $Q=1$ W, $D=10$ Kbits, and $T=5$ s.

\begin{figure}
\centering
\includegraphics[width=3in]{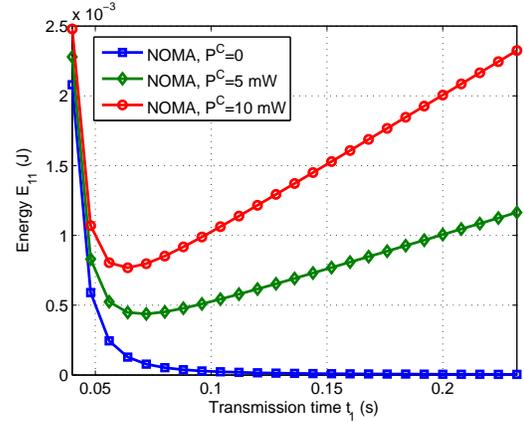}
\vspace{-1em}
\caption{Energy $E_{11}$ consumed by MTCD 1 versus the transmission time $t_1$ for NOMA strategy.}\label{fig5}
\vspace{-1em}
\end{figure}

Fig.~\ref{fig5} depicts, for instance, $E_{11}$ in (\ref{ee2eq3}) consumed by MTCD 1 served by MTCG 1 versus the transmission time $t_1$ for NOMA.
It is observed that $E_{11}$ monotonically decreases with $t_1$ when $P^C=0$.
For the case with $P^C=5$ mW or $P^C=10$ mW, $E_{11}$ first decreases and then increases with $t_1$.
Both observations validate our theoretical findings in Theorem 1.

\begin{figure}
\centering
\includegraphics[width=3in]{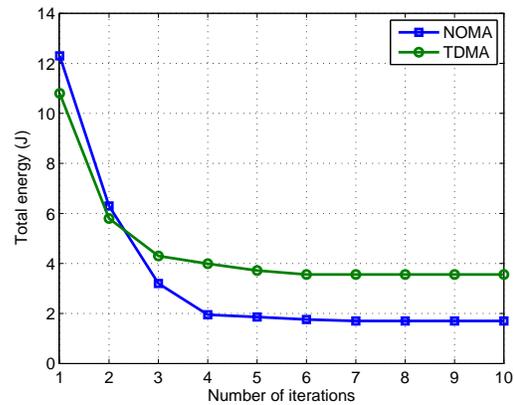}
\vspace{-1em}
\caption{Convergence behaviors of  IPCTA-NOMA and IPATC-TDMA.}\label{fig12}
\vspace{-1em}
\end{figure}

The convergence behaviors of IPCTA-NOMA and IPCTA-TDMA are illustrated in Fig.~\ref{fig12}.
From this figure, the total energy of both algorithms monotonically decreases, which confirms the convergence analysis in Section III-C and IV-C.
It can be seen that both IPCTA-NOMA and IPCTA-TDMA converge rapidly.

\begin{figure}
\centering
\includegraphics[width=3in]{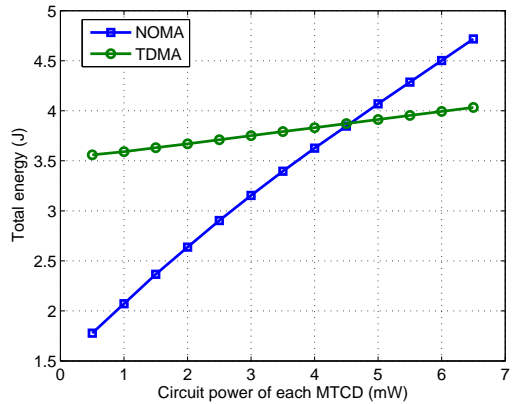}
\vspace{-1em}
\caption{Total energy versus the circuit power of each MTCD.}\label{fig10}
\vspace{-1em}
\end{figure}

In Fig.~\ref{fig10}, we illustrate the total energy consumption versus the circuit power of each MTCD.
According to Fig.~\ref{fig10}, the total energy of NOMA outperforms TDMA when the circuit power of each MTCD is low, i.e., $P^C\leq4$ mW in the test case.
At low circuit power regime, the total energy consumption of the network mainly lies in the RF transmission power of MTCDs and the energy consumed by MTCGs to charge the MTCDs through EH.
For the NOMA strategy, MTCDs served by the same MTCG can simultaneously upload data and the MTCG decodes the messages according to NOMA detections, which requires lower RF transmission power of MTCDs than the TDMA strategy.
Thus, the total energy of NOMA is less than TDMA for low circuit power of each MTCD. From Fig.~\ref{fig10}, we can find that the total energy of TDMA outperforms NOMA when the circuit power of each MTCD becomes high, i.e., $P^C\geq5$ mW in our tests.
At high circuit power regime, the total energy consumption of the network mainly lies in the circuit power of MTCDs and the energy consumed by MTCGs to charge the MTCDs through EH.
For the NOMA strategy, the transmission time of each MTCD with NOMA is always longer than that with TDMA, which leads to higher circuit power consumption of MTCDs than the TDMA strategy.
As a result, TDMA enjoys better energy efficiency than NOMA for high circuit power of each MTCD.

\begin{figure}
\centering
\includegraphics[width=3in]{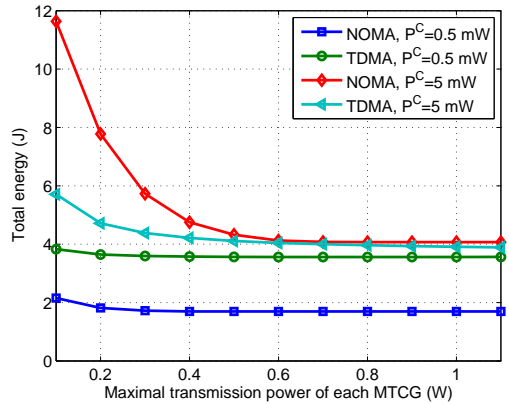}
\vspace{-1em}
\caption{Total energy versus the maximal transmission power of each MTCG.}\label{fig6}
\vspace{-1em}
\end{figure}

Fig.~\ref{fig6}  illustrates the total energy versus the maximal transmission power of each MTCG.
It is observed that the total energy decreases with the increase of maximal transmission power of each MTCG for both NOMA and TDMA.
This is because that the increment of maximal transmission power of each MTCG allows the MTCG to transmit with large transmission power, which leads to short EH time of MTCDs to harvest enough energy and low total energy consumption of the network.
Moreover, it is found that the total energy of NOMA is more sensitive to the maximal transmission power of each MTCG than that of TDMA for high circuit power case as $P^{\text C}=5$ mW for each MTCD.
The reason is that MTCD with low channel gain receives intra-cluster interference due to NOMA and the energy consumption is hence especially large for low maximal transmission power of each MTCG and high circuit power of each MTCD.

\begin{figure}
\centering
\includegraphics[width=3in]{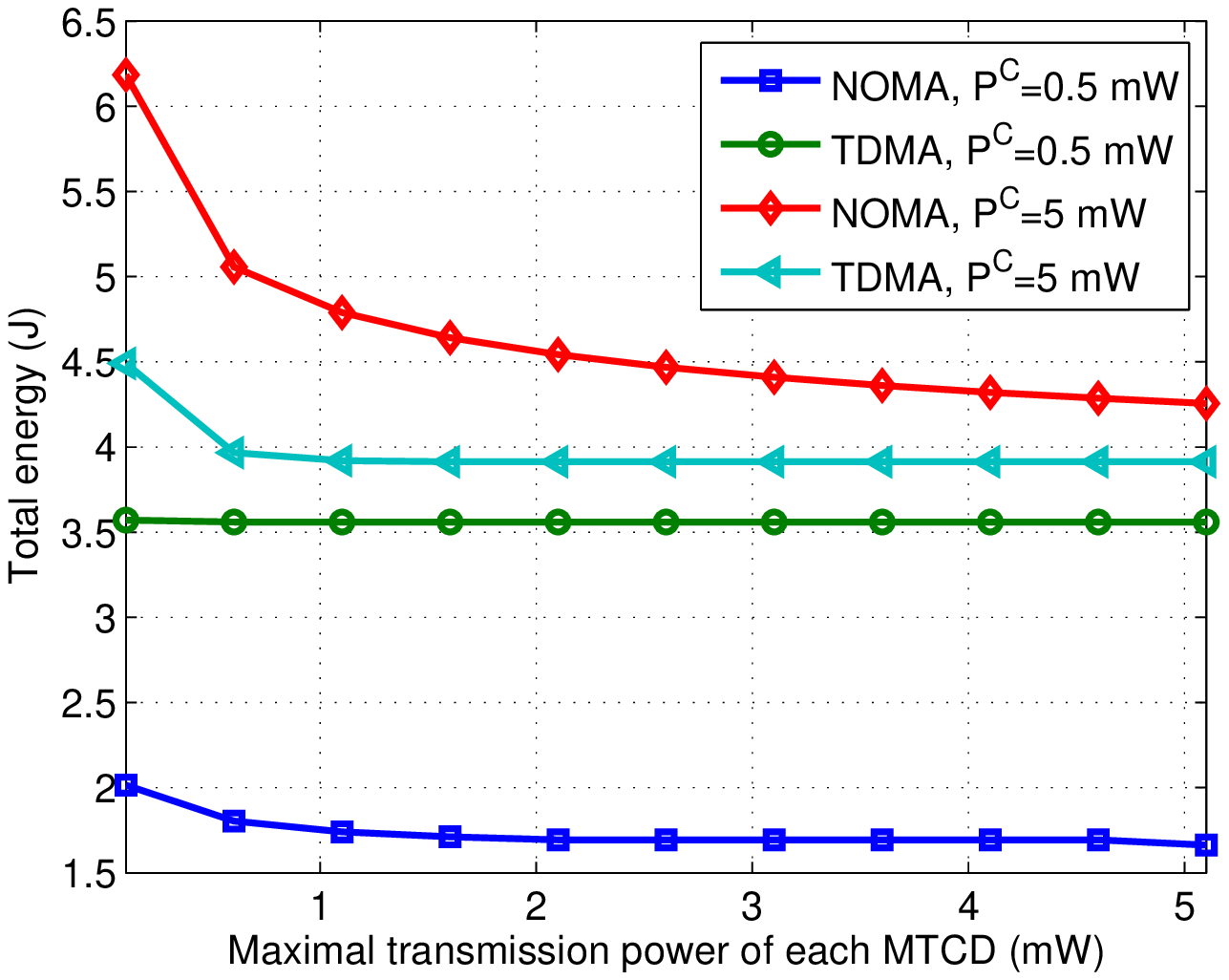}
\vspace{-1em}
\caption{Total energy versus the maximal transmission power of each MTCD.}\label{fig9}
\vspace{-1em}
\end{figure}

The total energy versus the maximal transmission power of each MTCD is shown in Fig.~\ref{fig9}.
It is observed that the total energy decreases with growing maximal transmission power of each MTCD for both NOMA and TDMA.
This is due to the fact that a larger maximal transmission power of each MTCD ensures MTCDs can transmit with more power,
and the required payload can be uploaded in a shorter time,
which results in low circuit power consumption and low energy consumption.
It can be found that the total energy of TDMA converges faster than that of NOMA as the maximal transmission power of each MTCD increases.
This is because that the MTCDs served by the same MTCG simultaneously transmit data for NOMA, and the required transmission power of each MTCD for NOMA is always larger than that of MTCD for TDMA.

%Fig.~\ref{fig6} and Fig.~\ref{fig9} illustrate the total energy versus the maximal transmission power of each MTCG and MTCD, respectively.
%It is observed that the total energy decreases with the increase of maximal transmission power of each MTCG and MTCD for both NOMA and TDMA.
%This is because the increase of maximal transmission power of each MTCG or MTCD allows the MTCG or MTCD to transmit with large transmission power, which results in both small transmission time and low energy consumption.

\begin{figure}
\centering
\includegraphics[width=3in]{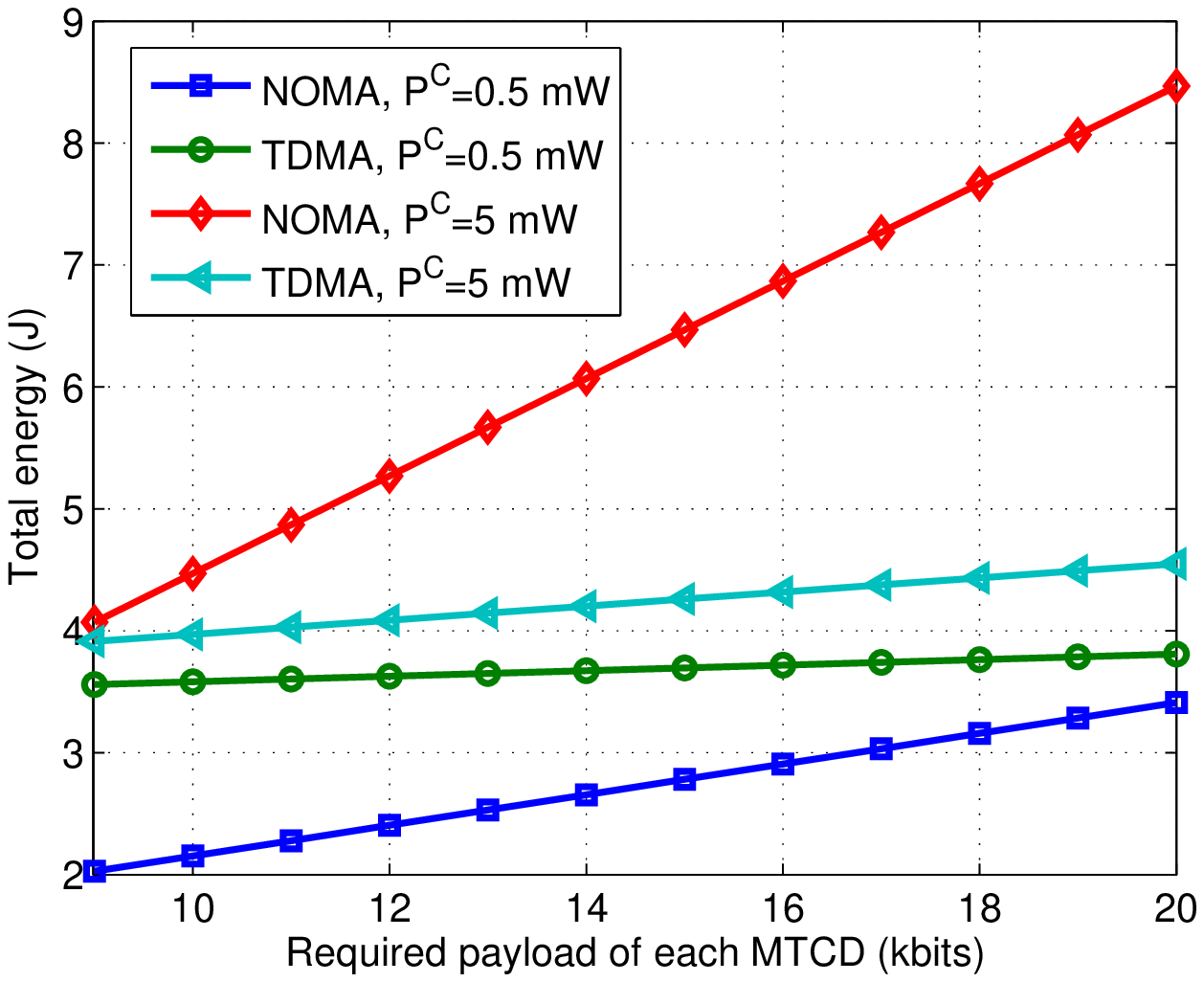}
\vspace{-1em}
\caption{Total energy versus the required payload of each MTCD.}\label{fig7}
\vspace{-1em}
\end{figure}

Finally, in Fig.~\ref{fig7}, we illustrate the total energy versus the required payload of each MTCD.
The figure shows that the total energy increases with the required payload of each MTCD.
This is due to the fact that large payload of each MTCD requires large energy consumption of MTCDs and MTCGs, which leads to high energy consumption of the network.

\section{Conclusions}
This paper compares the total energy consumption between NOMA and TDMA strategies in uplink M2M communications with EH.
We formulate the total energy minimization problem subject to minimal throughput constraints, maximal transmission power constraints and energy causality constraints, with the circuit power consumption taken into account.
By applying the conditions that it is optimal to transmit with the minimal throughput for each MTCD, we transform the original problem for NOMA strategy into an equivalent problem, which is suboptimally solved through an iterative algorithm.
By using a proper variable transformation, we transform the nonconvex problem for TDMA into an equivalent problem, which can be effectively solved.
Through simulations, either NOMA strategy or TDMA strategy may be preferred depending on different circuit power regimes of MTCDs.
At low circuit power regime of MTCDs, NOMA consumes less energy, while TDMA is preferred at high circuit power regime of MTCDs since the energy consumption for NOMA increases significantly as the circuit power of MTCDs increases.

\appendices
\section{Proof of Lemma 2}
\setcounter{equation}{0}
\renewcommand{\theequation}{\thesection.\arabic{equation}}

By inserting $r_{ij}=D_j$ into (\ref{sys1eq3}) from Lemma 1, we have
\begin{equation}\label{appenAeq1}
2^{\frac{D_{j}}{Bt_i}} \sum_{l=j+1}^{J_i} |h_{il}|^2p_l
+
\sigma^2 \left(2^{\frac{D_{j}}{Bt_i}} -1\right)
=  \sum_{l=j}^{J_i} |h_{il}|^2 p_l,
\end{equation}
for $j=J_{i-1}+1, \cdots, J_i$.
%Denoting $\pmb p_i=[p_{J_{i-1}+1}, \cdots, p_{J_i}]^T$, we can rewrite those $J_i-J_{i-1}$ equations in (\ref{appenAeq1}) as
%\begin{equation}\label{appenAeq1_2}
%E=\begin{bmatrix}
%1&-a_1\\
%&1 &-a_2 & \text{{\huge 0}}\\
%&&\ddots&\ddots\\
%& \text{{\huge 0}} && 1&-a_{m}\\
%&&&& 1
%\end{bmatrix}
%\end{equation}
To solve those $J_i-J_{i-1}$ equations, we first define
\begin{equation}\label{appenAeq2}
u_{j}=\sum_{l=j}^{J_i} |h_{il}|^2 p_{l}, \quad \forall j\in\mathcal  J_i.
\end{equation}
Applying (\ref{appenAeq2}) into (\ref{appenAeq1}) yields
\begin{equation}\label{appenAeq2_1}
{u_{j}}={2^{\frac{D_{j}}{Bt_i}}} u_{{j+1}}+
 {\sigma^2 }
 \left(2^{\frac{D_{j}}{Bt_i}} -1\right), \quad \forall j\in \mathcal J_i .
\end{equation}
Denote $\pmb u_i=[u_{J_{i-1}+1}, \cdots, u_{J_i}]^T$,
\begin{equation}\label{appenAeq2_2_0}
\pmb v_i=\left[{\sigma^2 }
 \left(2^{\frac{D_{J_{i-1}+1}}{Bt_i}} -1\right), \cdots, {\sigma^2 }
 \left(2^{\frac{D_{J_i}}{Bt_i}} -1\right)\right]^T,
\end{equation}
and
\begin{equation}\label{appenAeq2_2_2}
\pmb W_i=\begin{bmatrix}
0&{2^{\frac{D_{J_{i-1}+1}}{Bt_i}}}\\
&0 &\!\!\!\!\!\!\!\!{2^{\frac{D_{J_{i-1}+2}}{Bt_i}}} &  \\
&&\ddots&\ddots\\
&   && 0&{2^{\frac{D_{J_{i}-1}}{Bt_i}}}\\
&&&& 0
\end{bmatrix}.
\end{equation}
Equations in (\ref{appenAeq2_1}) can be rewritten as
\begin{equation}\label{appenAeq2_2_5}
(\pmb E - \pmb W_i) \pmb u_i=\pmb v_i,
\end{equation}
where $\pmb E$ is an identity matrix of size $(J_i-J_{i-1})\times(J_i-J_{i-1})$.
From (\ref{appenAeq2_2_5}), we have
\begin{equation}\label{appenAeq2_2_6}
\pmb u_i=(\pmb E - \pmb W_i)^{-1}\pmb v_i.
\end{equation}
Before obtaining the inverse matrix $(\pmb E - \pmb W_i)^{-1}$, we present the following lemma.
\begin{lemma}
When $l\in[1,J_i-J_{i-1}-1]$, the $l$-th power of matrix $\pmb W_i$ can be expressed as
\begin{equation}\label{appenAeq2_2_7}
\pmb W_i^l\!=\!\begin{bmatrix}
\pmb 0_{l}^T&\!\!\!\!{2^{\frac{\sum_{s=J_{i-1}+1}^{J_{i-1}+l}\!D_{s}}{Bt_i}}}\!\!\\
&\!\!\!\!0 &\!\!\!\!\!\!\!\!\!\!\!\!{2^{\frac{\sum_{s=J_{i-1}+2}^{J_{i-1}+l+1}\!D_{s}}{Bt_i}}} \!\!&  \\
&&\!\!\!\!\!\!\!\!\!\!\!\!\!\!\!\!\!\!\!\!\ddots&\!\!\!\!\!\!\!\!\ddots\\
&   && \!\!\!\!0&\:\; {2^{\frac{\sum_{s=J_{i}-l}^{J_{i}-1}\!D_{s}}{Bt_i}}}\!\!\\
&&&& \!\!\!\!\pmb 0_l\!\!
\end{bmatrix},
\end{equation}
where $\pmb 0_{l}$ denotes a $l\times 1$ vector of zeros.
When $l= J_i - J_{i-1}$, $\pmb W_i^l=\pmb 0_{(J_i-J_{i-1})\times(J_i-J_{i-1})}$, where
$\pmb 0_{(J_i-J_{i-1})\times(J_i-J_{i-1})}$ is a ${(J_i-J_{i-1})\times(J_i-J_{i-1})}$ matrix of zeros.
\end{lemma}

\itshape \textbf{Proof:} \upshape Lemma 4 can be proved by the principle of mathematical induction.

\emph{Basis}: It can be verified that Lemma 4 is valid for $l=1$.

\emph{Induction Hypothesis}: For $l\in[1,J_i-J_{i-1}-2]$, assume that the $l$-th power of matrix $\pmb W_i$ can be expressed as (\ref{appenAeq2_2_7}).

\emph{Induction Step}: According to (\ref{appenAeq2_2_7}), we can obtain
\begin{eqnarray}
&&\!\!\!\!\!\!\!\!\!\!\pmb W_i^{l+1}\!
=\!\pmb W_i^{l} \pmb W_i
\nonumber\\
&&\!\!\!\!\!\!\!\!\!\!
=
\begin{bmatrix}
\pmb 0_{l}^T&\!\!\!\!{2^{\frac{\sum_{s=J_{i-1}+1}^{J_{i-1}+l}\!D_{s}}{Bt_i}}}\!\!\\
&\!\!\!\!0 &\!\!\!\!\!\!\!\!\!\!\!\!{2^{\frac{\sum_{s=J_{i-1}+2}^{J_{i-1}+l+1}\!D_{s}}{Bt_i}}} \!\!&  \\
&&\!\!\!\!\!\!\!\!\!\!\!\!\!\!\!\!\!\!\!\!\ddots&\!\!\!\!\!\!\!\!\ddots\\
&   && \!\!\!\!0&\:\; {2^{\frac{\sum_{s=J_{i}-l}^{J_{i}-1}\!D_{s}}{B t_i}}}\!\!\\
&&&& \!\!\!\!\pmb 0_l\!\!
\end{bmatrix}
\nonumber\\
&&\!\!\!\!\!\!\!\!\!\!
\quad
\cdot
\begin{bmatrix}
0&{2^{\frac{D_{J_{i-1}+1}}{B t_i}}}\\
&0 &\!\!\!\!\!\!\!\!{2^{\frac{D_{J_{i-1}+2}}{B t_i}}} &  \\
&&\ddots&\ddots\\
&   && 0&{2^{\frac{D_{J_{i}-1}}{B t_i}}}\\
&&&& 0
\end{bmatrix}
\nonumber\\
&&\!\!\!\!\!\!\!\!\!\!
=
\begin{bmatrix}
\pmb 0_{l+1}^T&\!\!\!\!{2^{\frac{\sum_{s=J_{i-1}+1}^{J_{i-1}+l+1}\!D_{s}}{B t_i}}}\!\!\\
&\!\!\!\!0 &\!\!\!\!\!\!\!\!\!\!\!\!{2^{\frac{\sum_{s=J_{i-1}+2}^{J_{i-1}+l+2}\!D_{s}}{B t_i}}} \!\!&  \\
&&\!\!\!\!\!\!\!\!\!\!\!\!\!\!\!\!\!\!\!\!\ddots&\!\!\!\!\!\!\!\!\ddots\\
&   && \!\!\!\!0&\:\;{2^{\frac{\sum_{s=J_{i}-l-1}^{J_{i}-1}\!D_{s}}{B t_i}}}\!\!\\
&&&& \!\!\!\!\pmb 0_{l+1}\!\!
\end{bmatrix},\nonumber
\end{eqnarray}
which verifies that the $(l+1)$-th power of matrix $\pmb W_i$ can also be expressed as (\ref{appenAeq2_2_7}).
According to (\ref{appenAeq2_2_7}) and (\ref{appenAeq2_2_2}), it is verified that
\begin{equation}\label{appenAeq2_2_8}
\pmb W_i^{J_i-J_{i-1}}=\pmb W_i^{J_i-J_{i-1}-1}\pmb W_i=\pmb 0_{(J_i-J_{i-1})\times (J_i-J_{i-1})}.
\end{equation}
Therefore, Lemma 4 is proved.
\hfill $\Box$

Now, it is ready to obtain the inverse matrix $(\pmb E - \pmb W_i)^{-1}$.
Since
\begin{equation}\label{appenAeq2_2_9}
(\pmb E - \pmb W_i)\left(\pmb E +\sum_{l=1}^{{J_i-J_{i-1}-1}} \pmb W_i^l\right)=
\pmb E - \pmb W_i^{J_i-J_{i-1}}=\pmb E,
\end{equation}
we have
\begin{equation}\label{appenAeq2_2_9_1}
(\pmb E - \pmb W_i)^{-1}=\pmb E +\sum_{l=1}^{{J_i-J_{i-1}-1}} \pmb W_i^l.
\end{equation}

Substituting (\ref{appenAeq2_2_7}) and (\ref{appenAeq2_2_9_1}) into (\ref{appenAeq2_2_6}) yields
\begin{equation}\label{appenAeq2_2_9_2}
u_j=\sum_{l=j}^{J_i}
{\sigma^2}
\left(2^{\frac{D_{l}}{Bt_i}} -1\right) {2^{\frac{\sum_{s=j}^{l-1}D_{s}}{Bt_i}}}, \quad \forall  j\in \mathcal J_i,
\end{equation}
where we define $\prod_{s=j}^{j-1} {2^{\frac{D_{s}}{B t_i}}}=2^0$.
From (\ref{appenAeq2_2_9_2}), we can obtain $u_j={\sigma^2}\left({2^{\frac{\sum_{s=j}^{J_i}D_{s}}{Bt_i}}}-1\right)$.
Since $\sum_{l=J_i+1}^{J_i}p_{j}=0$, we have
\begin{equation}\label{appenAeq2_2}
u_{J_i+1}=0.
\end{equation}
%Based on (\ref{appenAeq2_1}) and (\ref{appenAeq2_2}), we have
%\begin{equation}\label{appenAeq2_3}
%b_{j}\!=\!\sum_{l=j}^{J_i}
%{\sigma^2}
%\left(2^{\frac{D_{l}}{Bt_i}} -1\right) {2^{\sum_{s=J_i+j-l}^{J_i-1}\frac{D_{s}}{Bt_i}}}, \quad \forall  j\in \mathcal J_i,\!\!\!
%\end{equation}
%where we define $\prod_{s=J_i}^{J_i-1} {2^{\frac{D_{s}}{B t_i}}}=1$.
From (\ref{appenAeq2}) and (\ref{appenAeq2_2}), we can obtain the transmission power of MTCD $j$ as
\begin{equation}\label{appenAeqpowervalue}
p_{ij}=\frac{u_{j}-u_{j+1}}
{|h_{ij}|^2},  \quad \forall j \in \mathcal J_i.
\end{equation}
Applying (\ref{appenAeq2_2_9_2}) and (\ref{appenAeq2_2}) to (\ref{appenAeqpowervalue}), we have
%(\ref{ee2eq2}).
%To show that $p_j$ decreases with $t_i$, equation (\ref{ee2eq2}) can be rewritten as
\begin{eqnarray}\label{appenAeq3}
 p_{j} &&\!\!\!\!\!\!\!\!\!\!
=\sum_{l=j}^{J_i}
\frac{ \sigma^2}
{|h_{ij}|^2 }
 \left(2^{\frac{D_{l}}{B t_i}} -1\right){2^{\frac{\sum_{s=j}^{l-1}D_{s}}{B t_i}}}
\nonumber \\
 &&\!\!\!\!\!\!\!\!\!\!\quad-
\sum_{l=j+1}^{J_i}
\frac{ \sigma^2}
{|h_{ij}|^2 }
\left(2^{\frac{D_{l}}{B t_i}} -1\right) {2^{\frac{\sum_{s=j+1}^{l-1}D_{s}}{B t_i}}}
\nonumber\\
&&\!\!\!\!\!\!\!\!\!\!
=
\!\sum_{l=j+1}^{J_i}
\frac{\sigma^2}
{|h_{ij}|^2 }
\!\left(\!2^{\frac{D_{l}}{B t_i}} \!-\!1\!\right)\! \left(\!{2^{\frac{D_{j}}{B t_i}}}\!-\!1\!\right)\!{2^{\frac{\sum_{s=j+1}^{l-1}D_{s}}{B t_i}}}
\nonumber\\
&&\!\!\!\!\!\!\!\!\!\!\quad+
\frac{\sigma^2}
{|h_{ij}|^2 }
\left(2^{\frac{D_{j}}{B t_i}} -1\right){2^{\frac{\sum_{s=j}^{j-1}D_{s}}{B t_i}}},
\nonumber\\
&&\!\!\!\!\!\!\!\!\!\!
=
 \sum_{l=j+1}^{J_i}
\frac{\sigma^2}
{|h_{ij}|^2 }
 \left( {\text e}^{\frac{a_{l}}{t_i}} - 1 \right) \left( {\text e^{\frac{a_{j}}{t_i}}} -  1 \right) {\text e^{ \frac{b_{jl}}{ t_i}}}
\nonumber\\
&&\!\!\!\!\!\!\!\!\!\!\quad+
\frac{\sigma^2}
{|h_{ij}|^2 }
\left(\text e^{\frac{a_{j}}{t_i}} -1\right),
\quad \forall j \in \mathcal J_i.
\end{eqnarray}
where $a_l$ and
$b_{jl}$ are defined in (\ref{ee2eq2_2}).
Since $\text{e}^{x} -1$ with $x\geq 0$ is non-negative and decreases with $t_i$,
$ p_{j}$ is also non-negative and decreases with $t_i$ from (\ref{appenAeq3}).

\section{Proof of Theorem 1}
\setcounter{equation}{0}
\renewcommand{\theequation}{\thesection.\arabic{equation}}
To show that the energy $E_{ij}$  is convex w.r.t. $t_i$, we first define function
\begin{eqnarray}\label{appenBeq1}
f_{ijl}(x)= { ( {\text e}^{ {a_{l}x} } - 1 )
 ( {\text e^{ {a_{j}x} }} -  1 )
\text e^{  {b_{jl}x} }}, \quad \forall x\geq 0.
\end{eqnarray}
Then, the second-order derivative follows
\begin{eqnarray}\label{appenBeq2}
f_{ijl}''(x)= &&\!\!\!\!\!\!\!\!\!\!
(a_{l}^2+2a_{l}b_{jl}) ( {\text e^{ {a_{j}x} }} - 1 )\text e^{ ({a_{l}+b_{jl}})x}
\nonumber\\
&&\!\!\!\!\!\!\!\!\!\!
+2a_{l} a_{j}  \text e^{ ({a_{l}+ a_{j}+b_{jl}})x}
\nonumber\\
&&\!\!\!\!\!\!\!\!\!\!
+(a_{j}^2+2a_{j}b_{jl})  ( {\text a^{ {a_{l}x} }} - 1 )\text e^{ ({a_{j}+b_{jl}})x}
\nonumber\\
&&\!\!\!\!\!\!\!\!\!\!
+b_{jl}^2
( {\text e}^{ {a_{l}x} } - 1 )
( {\text e^{ {a_{j}x} }} - 1 )\text e^{  b_{jl} x}\geq 0,
\end{eqnarray}
which indicates that $f_{ijl}(x)$ is convex w.r.t. $x$.
According to \cite[Page~89]{boyd2004convex}, the perspective of $u(x)$ is the function $v(x,t)$ defined by $v(x,t)=tu(x/t)$, ${\textbf{dom}}\:v= \{(x,t)|x/t\in {\textbf{dom}}\:u, t>0\}$.
If $u(x)$ is a convex function, then so is its perspective function $v(x,t)$ \cite[Page~89]{boyd2004convex}.
Since $f_{ijl}(x)$ is convex, the perspective function
\begin{equation}\label{appenBeq3}
\bar f_{ijl}(x, t_i)=t_i f_{ijl}\left(\frac x {t_i} \right)
\end{equation}
is convex w.r.t. ($x, t_i$).
Thus, function $\bar f_{ijl}(1, t_i)$ is also convex w.r.t. $t_i$.

Defining function
\begin{eqnarray}\label{appenBeq5_1}
g_{ij}(x)=
{\text e^{ {a_{j}}x} -1}, \quad
\end{eqnarray}
%we can obtain the second-order derivative as
%\begin{equation}\label{appenBeq5_2}
%g_{ij}''(x) =
%(a_{j}^2+2a_{j}d_{i}) \text e^{ ({a_{l}+d_{i}})x}
%+d_{i}^2 \text
%( {\text e}^{ {a_{l}x} } - 1 ) e^{  d_{i} x}
% \geq 0,
%\end{equation}
which is convex w.r.t. $x$.
By using the property of perspective function \cite[Page~89]{boyd2004convex}, we also have that function
\begin{equation}\label{appenBeq5_3}
\bar g_{ij}(x, t_i)=t_i g_{ij}\left(\frac x {t_i} \right)
\end{equation}
is convex  w.r.t. ($x, t_i$) and function $\bar g_{ij}(1, t_i)$ is accordingly convex w.r.t. $t_i$.

Substituting (\ref{appenBeq3}) and (\ref{appenBeq5_3}) into (\ref{ee2eq3}), we can obtain
\begin{equation}\label{appenBeq5_5}
E_{ij}=
\sum_{l=j+1}^{J_i}
\frac{\sigma^2}
{\eta |h_{ij}|^2 }
\bar f_{ijl}(1, t_i)
% \nonumber\\
%&&\!\!\!\!\!\!\!\!\!\! \quad
+
\frac{\sigma^2}
{\eta |h_{ij}|^2 }\bar g_{ij}(1, t_i)
+ t_i P_{\text C}.
\end{equation}
Due to the fact that both $\bar f_{ijl}(1, t_i)$ and $\bar g_{ij}(1, t_i)$ are convex,
$E_{ij}$ is consequently convex w.r.t. $t_i$ from (\ref{appenBeq5_5}).

According to (\ref{ee2eq3}), the first-order derivative of $E_i$ w.r.t. $t_i$ is expressed as
\begin{eqnarray}\label{appenBeq5_7}
\frac {\partial E_{ij}}{\partial t_i}&&\!\!\!\!\!\!\!\!\!\!=
\sum_{l=j+1}^{J_i}
\frac{\sigma^2 }
{\eta |h_{ij}|^2 }
{\left( {\text e}^{\frac{a_{l}}{t_i}} - 1 \right)
 \left( {\text e^{\frac{a_{j}}{t_i}}} -  1 \right)
\text e^{ \frac{b_{jl}}{  t_i}}}
 \nonumber\\
 &&\!\!\!\!\!\!\!\!\!\!\quad -
\sum_{l=j+1}^{J_i}
\frac{\sigma^2 a_l}
{\eta |h_{ij}|^2 t_i}
{
 \left( {\text e^{\frac{a_{j}}{t_i}}} -  1 \right)
\text e^{ \frac{a_l+b_{jl}}{  t_i}}}
 \nonumber\\
 &&\!\!\!\!\!\!\!\!\!\! \quad-
\sum_{l=j+1}^{J_i}
\frac{\sigma^2 a_{j}}
{\eta |h_{ij}|^2 t_i}
{
 \left( {\text e^{\frac{a_{l}}{t_i}}} - 1 \right)
\text e^{ \frac{a_{j}+b_{jl}}{  t_i}}}
 \nonumber\\
 &&\!\!\!\!\!\!\!\!\!\!\quad -
\sum_{l=j+1}^{J_i}
\frac{\sigma^2 b_{jl}}
{\eta |h_{ij}|^2 t_i }
{\left( {\text e}^{\frac{a_{l}}{t_i}} - 1 \right)
 \left( {\text e^{\frac{a_{j}}{t_i}}} - 1 \right)
\text e^{ \frac{b_{jl}}{  t_i}}}
 \nonumber\\
 &&\!\!\!\!\!\!\!\!\!\! \quad+
\frac{\sigma^2 }
{\eta |h_{ij}|^2 }
{\left(\text e^{\frac{a_{j}}{t_i}} -1\right)
}
% \nonumber\\
% &&\!\!\!\!\!\!\!\!\!\!
 -
\frac{\sigma^2 a_j}
{\eta |h_{ij}|^2 t_i}
{
\text e^{ \frac{a_j}{ t_i}}}+ P_{\text C}.
\end{eqnarray}
Since $E_{ij}$ is convex w.r.t. $t_i$, function $\frac{\partial E_{ij}}{\partial t_i}$ increases with $t_i$.
Because $D_j$ is positive for all $j$, we have $a_{l}>0$, $b_{ijl}>0$ and $c_{ijl}>0$ from (\ref{ee2eq2_2}).
To calculate the limit of $\frac{\partial E_{ij}}{\partial t_i}$ at $t_i=0+$, we calculate the following limit,
\begin{eqnarray}\label{appenBeq5_8}
&&\!\!\!\!\!\!\!\!\!\!\quad\lim_{t_i\rightarrow 0+}
 \frac{\sigma^2 }
{\eta |h_{ij}|^2 }
{\left( {\text e}^{\frac{a_{l}}{t_i}} - 1 \right)
 \left( {\text e^{\frac{a_{j}}{t_i}}} - 1 \right)
\text e^{ \frac{b_{jl}}{  t_i}}}
 \nonumber\\
 &&\!\!\!\!\!\!\!\!\!\! \qquad-
\frac{\sigma^2 a_l}
{\eta |h_{ij}|^2 t_i}
{
 \left( {\text e^{\frac{a_{j}}{t_i}}} - 1 \right)
\text e^{ \frac{a_l+b_{jl}}{  t_i}}}
\nonumber\\
&&\!\!\!\!\!\!\!\!\!\!=\lim_{x\rightarrow  +\infty}
 \frac{\sigma^2 }
{\eta |h_{ij}|^2 }
{\left( {\text e}^{ {a_{l}}x} - 1 \right)
 \left( {\text e^{ {a_{j}}x}} - 1 \right)
\text e^{  {b_{jl}}x}}
 \nonumber\\
 &&\!\!\!\!\!\!\!\!\!\! \qquad-
\frac{\sigma^2 a_l x}
{\eta |h_{ij}|^2 }
{
 \left( {\text e^{ {a_{j}}x}} - 1 \right)
\text e^{  ({a_l+b_{jl}})x}}
 \nonumber\\
 &&\!\!\!\!\!\!\!\!\!\!
=\lim_{x\rightarrow +\infty}
 \frac{\sigma^2 (1-a_l x)}
{\eta |h_{ij}|^2 }
{\text e}^{( {a_{l}}+{a_{j}}+b_{jl})x}
 \nonumber\\
 &&\!\!\!\!\!\!\!\!\!\!
=-\infty.
\end{eqnarray}
Thus, when $t_i$ approaches zero in the positive direction, we have
\begin{eqnarray}\label{appenBeq5_10}
\lim_{t_i \rightarrow 0+}\frac {\partial E_{ij}}{\partial t_i}
=-\infty.
\end{eqnarray}
When $t_i$ approaches positive infinity, the limit of first-order derivative $\frac {\partial E_{ij}}{\partial t_i}$ can be calculated as
\begin{eqnarray}\label{appenBeq5_11}
\lim_{t_i \rightarrow +\infty}\frac {\partial E_{ij}}{\partial t_i}
= P_{\text C}.
\end{eqnarray}

If $P_{\text C}=0$, then $\frac {\partial E_{ij}}{\partial t_i} \leq 0$ for all $t_i \geq 0$.
In this case, the energy $E_{ij}$ always decreases with $t_i$.
If $P_{\text C}>0$, we can observe that $\lim_{t_i \rightarrow +\infty}\frac {\partial E_{ij}}{\partial t_i}>0$ from (\ref{appenBeq5_11}).
Since $\lim_{t_i \rightarrow 0+}\frac {\partial E_{ij}}{\partial t_i}<0$ and $\frac{\partial E_{ij}}{\partial t_i}$ increases with $t_i$, there exists one unique solution $T_{ij}^*$ satisfying (\ref{ee2eq5}), which can be solved by using the bisection method.
In particular, $E_{ij}$ decreases with $0\leq t_i\leq T_{ij}^*$ and increases with $t_{ij} > T_{ij}^*$.

\section{Proof of Theorem 2}
\setcounter{equation}{0}
\renewcommand{\theequation}{\thesection.\arabic{equation}}
We first consider the case that $T\leq \max_{\forall i \in \mathcal N}$ $\min_{\forall j \in \mathcal J_i} \{T_{ij}^*\}$.
Without loss of generality, we denote $T\leq T_{nm}^*=\max_{\forall i \in \mathcal N}\min_{\forall j \in \mathcal J_i} \{T_{ij}^*\}$.
Assume that the optimal solution ($\pmb p^*, \pmb q^*, \pmb t^*$) to problem (\ref{ee2min1}) satisfies $\sum_{i=1}^{N+K}t_i^*<T$.
Due to that $t_i^*\geq0$ for all $i$, we can obtain that $t_n^*\leq T_{nm}^*\leq T_{nl}^*$, $\forall  l \in \mathcal J_n$.
With all other power $p_j^*$, $q_i^*$ and time $t_k^*$ fixed, $j\in\mathcal M\setminus\mathcal J_n$, $i\in \mathcal N$, $k\in\mathcal N\setminus\{n\}$,
we increase the time $t_n^*$ to $t_n'=t_n^*+\epsilon$ by an arbitrary amount $(0, T-\sum_{i=1,i\neq n}^N t_i^*]$.
Using (\ref{ee2eq2}), the corresponding power $p_l^*$ strictly deceases to $p_l'$, $l\in\mathcal J_n$.
According to Theorem 1, the energy $E_{nl}$ decreases with the transmission time $0\leq t_n \leq T_{nl}^*$, $\forall l \in \mathcal J_n$.
As a result, with new power-time pair ($p_{J_{n-1}+1}', \cdots, p_{J_n}', t_n'$), the objective function (\ref{ee2min1}a) decreases with all the constraints satisfied.
By contradiction, we must have $\sum_{i=1}^{N+K}t_i^*=T$ for the optimal solution.
This completes the proof of the first half part of Theorem 2.

%%%===================================================
The last half part of Theorem 2 indicates that transmitting with maximal transmission time is not optimal when $T$ is larger than a threshold $T_{\text{Upp}}$.
This can be proved by using the contradiction method.
Specifically, assuming that total transmission time of the optimal solution is the maximal transmission time $T$, we can find a special solution with total transmission time less than $T$, which strictly outperforms the optimal solution.
%We then prove the last half part of Theorem 2 by using the contradiction method, i.e., we show that there always exits a feasible solution with partial transmission time, which outperforms any feasible solution with full transmission time
%when the maximal transmission time $T$ exceeds a threshold.
% that transmitting with maximal transmission time is not optimal in minimizing the total energy consumption.
%According to Theorem 1, the optimal transmission time in minimizing the energy $E_{ij}$ consumed by MTCDs $j$ in $\mathcal J_i$ is $T_{ij}^*$ for all $i \in \mathcal N$.
%Thus, when the maximal transmission time $T$ is large enough, i.e., beyond a threshold,  the optimal transmission time is not the maximal transmission time.
To obtain the threshold $T_{\text{Upp}}$, we consider a special solution that satisfies all the constraints of problem (\ref{ee2min1}) except the maximal transmission time constraint (\ref{ee2min1}e).

%From Theorem 1, the energy $E_{ij}$ achieves the minimum at $t_i=T_{ij}^*$.
Since $E_{ij}$ is convex w.r.t. $t_i$ according to Theorem 1, the energy $E_{i}=\sum _{j =J_{i-1}+1}^{J_i} E_{ij}$ consumed by all MTCDs in $\mathcal J_i$ served by MTCG $i$ is also convex w.r.t. $t_i$.
Based on the proof of Theorem 1, we directly obtain the following lemma.
\begin{lemma}
The energy $E_i$ consumed by all MTCDs in $\mathcal J_i$ first decreases with the transmission time $t_i$ when $0\leq t_i \leq T_i^*$ and then increases with the transmission time $t_i$ when $t_i>T_i^*$, where $T_i^*$ is the unique zero point of first-order derivative $\frac {\partial E_i}{\partial t_i}$, i.e.,
\begin{eqnarray}\label{appenCee2eq5}
\left.\frac {\partial E_i}{\partial t_i}
\right|_{t_i=T_i^*}
=\sum _{j =J_{i-1}+1}^{J_i}\left.\frac {\partial E_{ij}}{\partial t_i}
\right|_{t_i=T_i^*}
=0.
\end{eqnarray}
\end{lemma}
Since Lemma 5 can be proved by checking the first-order derivative $\frac {\partial E_i}{\partial t_i}$ as in Appendix B, the proof of Lemma 5 is omitted here.

Set $\tilde t_i=T_i^*$, and $\tilde p_{j}$ can be obtained from (\ref{ee2eq2}) with $\tilde t_i$, $i \in \mathcal N$, $j \in \mathcal J_i$.
For the transmission power of the MTCGs, we set $\tilde q_i=Q_i$, $i \in \mathcal N$.
Denote $\tilde {\pmb p}=[\tilde p_1, \cdots, \tilde p_M]^T$,
$\tilde{\pmb q}=[\tilde q_1, \cdots, \tilde q_N]^T$.
With power $(\tilde {\pmb p}, \tilde {\pmb q})$ and time $(\tilde t_i, \cdots, \tilde t_N)$ fixed for now,
the energy minimization problem (\ref{ee2min1}) without constraint (\ref{ee2min1}e) becomes
\begin{subequations}\label{appenCee2min1}
\begin{align}
\!\!\mathop{\min}_{\pmb \tau}\;
\!& \sum_{k=1}^K  \sum_{i=I_{k-1}+1}^{I_k}\! t_{N+k} \!\left(\!\frac{Q_i}{\xi} \!+\! Q^{\text C}\!\right)
\nonumber\\
&
\!-\!\sum_{k=1}^K\sum_{i=1}^{ N}\! \sum_{j=J_{i-1}+1}^{ J_i}  \! t_{N+1}   u \left(\!\sum_{n=J_{k-1}+1}^{J_k} \!|h_{nj}|^2 Q_n \!\right)
\\
\textrm{s.t.}\qquad \!\!\!\!\!\!\!\!\!
&B t_{N+i}\log_2\!\left(\!
1\!+\!\frac{|h_{i}|^2 Q_i}
{ \sum_{n=i+1}^{I_k} \!|h_{n}|^2 Q_n\!+\!\sigma^2}
\!\right)\!
\!\geq \! \sum_{j\in\mathcal J_i} \!D_j,
\nonumber\\
&\qquad\qquad\forall k \in \mathcal K, i \in \mathcal I_k \\
&\tilde t_i\! \left(\!\frac{\tilde p_j}{\eta} \!+\! P^{\text C}\!\right)\!\leq\!
  \sum_{k=1}^Kt_{N+k} u\!\left(\!\sum_{n=I_{k-1}+1}^{I_k} \! |h_{nj}|^2 Q_n\!\right),
  \nonumber\\
  &\qquad\qquad
 \forall i \in \mathcal N, j \in \mathcal J_i
\\
&\pmb \tau \geq \pmb 0,
\end{align}
\end{subequations}
where $\pmb \tau=[t_{N+1}, \cdots, t_{N+K}]^T$.
Problem (\ref{appenCee2min1})
can be obtained by substituting (\ref{sys1eq8}) into (\ref{ee2min1}a), (\ref{sys1eq4}) into (\ref{ee2min1}b), and (\ref{sys1eq5}) and (\ref{sys1eq6}) into (\ref{ee2min1}c).
Problem (\ref{appenCee2min1}) is a linear problem, which can be optimally solved via the simplex method.
Denote the optimal solution of problem by $ \pmb \tau^*=[T_{N+1}^*, \cdots, T_{N+K}^*]^T$.
%According to (\ref{sys1eq5}) and (\ref{sys1eq7_2}), the objective function (\ref{appenCee2min1}a) monotonically increases with time $t_{N+1}$.
%Thus, the optimal solution to problem (\ref{appenCee2min1}) is the minimal $t_{N+1}$ satisfying all constraints (\ref{appenCee2min1}b)-(\ref{appenCee2min1}d).
%Denote the optimal solution to problem (\ref{appenCee2min1}) by $T_{N+1}^*$, which can be expressed as
%\begin{eqnarray}\label{appenCeqn1}
%T_{N+1}^*=&&\!\!\!\!\!\!\!\!\!\!
%\max\left\{
%\max_{i\in\mathcal N}\left\{\frac{\sum_{j\in\mathcal J_i}D_j}
%{B\log_2\!\left(\!
%1\!+\!\frac{|h_{i}|^2 Q_i}
%{ \sum_{n=i+1}^{N} \!|h_{n}|^2 Q_n\!+\!\sigma^2}
%\!\right)}\right\},\right.
%\nonumber\\
%&&\!\!\!\!\!\!\!\!\!\!
%\left.
%\qquad \max_{i\in \mathcal N, j\in \mathcal J_i}
%\left\{
%\frac{\tilde t_i \left(\frac{\tilde p_j}{\eta} + P^{\text C}\right)}
%{ u\left( \sum_{n=1}^N |h_{nj}|^2 Q_n \right)}
%\right\}
%\right\}.
%\end{eqnarray}
%%============================================0928
Denote $\tilde t_{N+k}=T_{N+k}^*$, $\forall k \in \mathcal K$, and $\tilde {\pmb t}=[\tilde t_{1}, \cdots, \tilde t_{N+K}]^T$.
%To ensure that time vector $\tilde {\pmb t}$ meets the maximal transmission time constraint (\ref{ee2min1}e),
%$T$ should satisfy
%\begin{equation}\label{appenCeqn2}
%T\geq \sum_{i=1}^N T_i^*.
%\end{equation}
As a result, we obtain a special solution ($\tilde {\pmb p}, \tilde {\pmb q}, \tilde {\pmb t}$) that satisfies all the constraints of problem (\ref{ee2min1}) except the maximal transmission time constraint (\ref{ee2min1}e).
With solution ($\tilde {\pmb p}, \tilde {\pmb q}, \tilde {\pmb t}$),
the total energy consumption $\tilde E_{\text {Tot}}$ obtained from (\ref{sys1eq8}) can be expressed as
\begin{eqnarray}\label{appenCeqn5}
\!\!\!\!\!\!\tilde E_{\text {Tot}}
&&\!\!\!\!\!\!\!\!\!\!=\sum_{i=1}^{N+K} \tilde E_i
\nonumber\\
&&\!\!\!\!\!\!\!\!\!\!=\sum_{i=1}^{ N}\sum_{j=J_{i-1}+1}^{ J_i}\tilde t_i \left(\frac{\tilde p_j}{\eta}
 + P^{\text C}\right)\nonumber\\
&&\!\!\!\!\!\!\!\!\!\!
\quad+\sum_{k=1}^K\sum_{i=I_{k-1}+1}^{I_k}\tilde  t_{N+k} \left(\frac{\tilde q_i}{\xi} + Q^{\text C}\right)
\nonumber\\
&&\!\!\!\!\!\!\!\!\!\!
\quad\!-\!\sum_{k=1}^K\sum_{i=1}^{ N}\! \sum_{j=J_{i-1}+1}^{ J_i} \!  \tilde t_{N+k} u\!\left(\!\sum_{n=I_{k-1}+1}^{I_k}\! |h_{nj}|^2\tilde q_n\!\right)\!\!,
\end{eqnarray}
where $\tilde E_i$ is the energy consumed by all MTCDs in $\mathcal J_i$, $i\in \mathcal N$, and $\tilde E_{N+k}$ is the system energy consumption during the ($N+k$)-th phase.

Denote an upper bound of maximal transmission time by
\begin{equation}\label{appenCeqn6}
T_{\text{Upp}}=\max\left\{\sum_{i=1}^{N+K} T_i^*,  T_{\text{Amp}}  \right\},
\end{equation}
where
\begin{equation}
T_{\text{Amp}}=
\frac{\left(
1+\sum_{i=1}^N\beta_i\right)\sum_{k=1}^K\tilde E_{N+k}
}
{\alpha Q^{\text C}},
\end{equation}
with $\alpha$ defined in (\ref{appenCeqn7_5}) and $\beta_i$ defined in (\ref{appenCeqn7_55}).
If $T\geq T_{ \text{Upp}}$, we show that optimal solution ($\pmb p^*, \pmb q^*, \pmb t^*$) to problem (\ref{ee2min1}) must satisfy constraint (\ref{ee2eq6_3}), i.e., (\ref{ee2min1}e) is inactive, by contradiction.
Assume that
\begin{equation}\label{appenCeqn6_2}
\sum_{i=1}^{N+K}t_i^*=T.
\end{equation}
With ($\pmb p^*, \pmb q^*, \pmb t^*$),
we denote $E_i^*$ as the energy consumed by all MTCDs in $\mathcal J_i$, $i\in \mathcal N$, $E_{N+k}^*$ as the system energy consumption during the ($N+k$)-th phase, and $E_{\text {Tot}}^*$ as the total energy of the whole system.
Thus, we have
\begin{eqnarray}\label{appenCeqn7}
E_{\text {Tot}}^*&&\!\!\!\!\!\!\!\!\!\!=\sum_{i=1}^{N+K} E_i^*
\nonumber\\
&&\!\!\!\!\!\!\!\!\!\!
\overset{(\text a)}{\geq}
\sum_{i=1}^{N+1} \tilde E_i+\sum_{k=1}^K E_{N+k}^*
\nonumber\\
&&\!\!\!\!\!\!\!\!\!\!
\overset{(\text b)}{=}
\sum_{i=1}^{N}\! \tilde E_i\! +\!
\sum_{k=1}^{K} \!  t_{N+k}^*  \left( \!\sum_{n=I_{k-1}+1}^{I_k}\frac{ q_n^*}{\xi}\!
-
\!\sum_{i=1}^{ N} \!\sum _{j=J_{i-1}+1}^{J_i}
 \right.
\nonumber\\
&&\!\!\!\!\!\!\!\!\!\!
\left.
\quad u\!\left(\!\sum_{n=I_{k-1}+1}^{ I_k}\!|h_{nj}|^2 q_n^* \!\right)\!\right)
%\nonumber\\
%&&\!\!\!\!\!\!\!\!\!\!
%\quad
+ Q^{\text C} \sum_{k=1}^K t_{N+k}^* (I_k-I_{k-1})
\nonumber\\
&&\!\!\!\!\!\!\!\!\!\!
\overset{(\text c)}{>}
\sum_{i=1}^{N} \tilde E_i+\alpha Q^{\text C} \sum_{k=1}^K t_{N+k}^*
\nonumber\\
&&\!\!\!\!\!\!\!\!\!\!\overset{(\text d)}{\geq}\sum_{i=1}^{N+K} \tilde E_i
=\tilde E_{\text {Tot}},
\end{eqnarray}
where inequality (a) follows from the fact that $E_i$ achieves the minimum when $t_i=T_i^*$ according to Lemma 5,
equality (b) holds from (\ref{sys1eq5}) and (\ref{sys1eq7}),
and inequality (c) follows from (\ref{sys1eq5}), (\ref{sys1eq7_2}), $\xi\in(0,1]$ and
\begin{equation}\label{appenCeqn7_5}
\alpha\triangleq\min_{k\in\mathcal K}(I_k-I_{k-1}).
\end{equation}
 %and inequality (d) follows from (\ref{appenCeqn6}).
To explain procedure (d),
we substitute (\ref{sys1eq5}) and (\ref{sys1eq6}) into energy causality constraints (\ref{ee2min1}d) to obtain
\begin{equation}\label{appenCeqn7_2}
t_i^* \left(\frac{p_j^*}{\eta} + P^{\text C}\right)\leq \sum_{k=1}^K  t_{N+k}^* u\left(\sum_{n=I_{k-1}+1}^{I_k} |h_{nj}|^2 q_n\right), %\quad \forall i \in\mathcal N, j \in \mathcal J_i.
\end{equation}
for all $i \in \mathcal N, j \in \mathcal J_i$.

Considering that $p_j^*\geq 0$ %from (\ref{ee2min1}f)
in the left hand side of (\ref{appenCeqn7_2}) and $u(x)$ is a increasing function as well as $q_i^*\leq Q_i$ in the right hand side of (\ref{appenCeqn7_2}), we have
\begin{eqnarray}\label{appenCeqn7_5_2}
t_i^*&&\!\!\!\!\!\!\!\!\!\! \leq\min_{j\in\mathcal J_i}\left\{\frac{ \sum_{k=1}^Kt_{N+k}^* u\left(\sum_{n=I_{k-1}+1}^{I_k}|h_{nj}|^2 Q_n \right)}{P^C}\right\},
\nonumber\\
&&\!\!\!\!\!\!\!\!\!\!
 \leq \beta_i\sum_{k=1}^Kt_{N+k}^*
, \quad\forall i\in \mathcal N,
\end{eqnarray}
where
\begin{equation}\label{appenCeqn7_55}
\beta_i=\min_{j\in\mathcal J_i}\left\{\max_{k\in\mathcal K}\frac{u\left(\sum_{n=I_{k-1}+1}^{I_k}|h_{nj}|^2 Q_n \right)}{P^C}\right\}.
\end{equation}
Combining (\ref{appenCeqn6_2}) and (\ref{appenCeqn7_5}) yields
\begin{equation}\label{appenCeqn7_6}
\sum_{k=1}^K t_{N+k}^*\geq \frac{T}
 {1+\sum_{i=1}^N\beta_i}.
\end{equation}
Hence, inequality (d) follows from (\ref{appenCeqn6}) and (\ref{appenCeqn7_6}).

According to (\ref{appenCee2min1}) and (\ref{appenCeqn6}), solution ($\tilde {\pmb p}, \tilde{\pmb q}, \tilde{\pmb t}$) is a feasible solution to problem (\ref{ee2min1}).
From (\ref{appenCeqn7}), the objective value (\ref{ee2min1}a) can be decreased with solution ($\tilde {\pmb p}, \tilde{\pmb q}, \tilde{\pmb t}$),
which contradicts that ($\pmb p^*, \pmb q^*, \pmb t^*$) is the optimal solution to problem (\ref{ee2min1}).
Hence, the optimal solution  to problem (\ref{ee2min1}) must satisfy constraint (\ref{ee2eq6_3}).

\section{Proof of Theorem 3}
\setcounter{equation}{0}
\renewcommand{\theequation}{\thesection.\arabic{equation}}
We first show that the feasible set of problem (\ref{ee2min2}) with given $\pmb \tau$ is convex.
Obviously, constraints (\ref{ee2min2}b), (\ref{ee2min2}d), (\ref{ee2min2}e) and (\ref{ee2min2}g) and (\ref{ee2min2}h) are all linear w.r.t. ($\pmb q, \bar {\pmb t}$).
According to (\ref{appenBeq1}) and (\ref{appenBeq3}), constraints (\ref{ee2min2}c) and (\ref{ee2min2}f) can be, respectively, reformulated as
\begin{equation}\label{appenDeq1}
\sum_{l=j+1}^{J_i}
\bar f_{ijl}(1, t_i)  + t_i P^{\text C}
 \leq  \sum_{k\in\mathcal S_{ij}}   t_{N+k} \bar u\left(\sum_{n=I_{k-1}+1}^{ I_k } |h_{nj}|^2 q_n \right)   %i\in \mathcal N, j \in \mathcal J_i,
\end{equation}
for all $i\in \mathcal N, j \in \mathcal J_i$,
and
\begin{equation}\label{appenDeq2}
\sum_{l=j+1}^{J_i}
 f_{ijl}\left(\frac 1 {t_i} \right)
 \leq  P_j, \quad\forall i\in \mathcal N, j \in \mathcal J_i.
\end{equation}

Based on (\ref{sys1eq5_22}), we have
\begin{equation}\label{appenDeq2_2}
\bar u''(x)=\frac{-M a^2 (1+\text e ^{ab}) \text e^{-a(x-b)}}{\text e^{ab} (1+\text{e} ^{-a(x-b)})^3}
\leq 0, \quad \forall x\geq 0,
\end{equation}
which shows that $\bar u(x)$ is concave, and $-\bar u(x)$ is convex.
Since $\bar f_{ijl}(1, t_i)$ is convex w.r.t. $t_i$ according to Appendix B, constraints (\ref{appenDeq1})  are convex, i.e., constraints (\ref{ee2min2}c) are convex.
Due to that $f_{ijl}(x)$ is convex and non-decreasing, and $\frac 1 x$ is convex, $f_{ijl}\left(\frac 1 {x} \right) $ is also convex according to the composition property of convex functions \cite[Page~84]{boyd2004convex}.
Thus, constraints (\ref{appenDeq2})  are convex, i.e., constraints (\ref{ee2min2}f) are convex.

We then show that the objective function (\ref{ee2min2}a) is convex.
Substituting (\ref{appenBeq3}) and (\ref{appenBeq5_3}) into (\ref{ee2min2}a) yields
\begin{eqnarray}\label{appenDeq3}
&&\!\!\!\!\!\!\!\!\!\!\quad\sum_{i=1}^N \sum_{j=J_{i-1}+1}^{ J_i}
\sum_{l=j+1}^{J_i}
\frac{\sigma^2}
{\eta |h_{ij}|^2 }
\bar f_{ijl}(1, t_i)
\nonumber\\
&&\!\!\!\!\!\!\!\!\!\! +\sum_{i=1}^N \sum_{j=J_{i-1}+1}^{ J_i}
 \frac{\sigma^2}
{\eta |h_{ij}|^2 }
\bar g_{ij}(1, t_i)
\nonumber\\
&&\!\!\!\!\!\!\!\!\!\!+\sum_{i=1}^N \sum_{j=J_{i-1}+1}^{ J_i}t_i P_{\text C}
+\sum_{k=1}^K\sum_{i=I_{k-1}+1}^{I_k} t_{N+k} \left(\frac{q_i}{\xi} + Q^{\text C}\right)
\nonumber\\
&&\!\!\!\!\!\!\!\!\!\! -\sum_{i=1}^{ N} \sum_{j=J_{i-1}+1}^{ J_i} \sum_{k\in\mathcal S_{ij}} t_{N+k} \bar u\left(\sum_{n=I_{k-1}+1}^{ I_k} |h_{nj}|^2 q_n\right),
\end{eqnarray}
which is convex w.r.t. ($\pmb q, \bar{\pmb t}$) because $\bar f_{ijl}(1, t_i)$, $\bar g_{ij}(1, t_i)$ and $-\bar u(x)$ are convex according to Appendix B and (\ref{appenDeq2_2}).
As a result, problem (\ref{ee2min2}) is convex w.r.t. ($\pmb q, \bar {\pmb t}$).

With given ($\pmb q, \bar {\pmb t}$), constraints (\ref{ee2min2}b) can be equivalently transformed into
\begin{equation}\label{appendeq5}
B t_{N+i}\log_2\left(
1+\frac{|h_{i}|^2 q_i}
{ \sum_{n=i+1}^{I_k} |h_{n}|^2 q_n+\sigma^2}
\right) \geq \sum_{j=J_{i-1}+1}^{ J_i} D_j,
\end{equation}
which is linear w.r.t. $t_{N+i}$.
By replacing constraints (\ref{ee2min2}b) with (\ref{appendeq5}),
problem (\ref{ee2min2}) with given ($\pmb q, \bar {\pmb t}$) is a linear problem.

\section{Proof of Theorem 4}
\setcounter{equation}{0}
\renewcommand{\theequation}{\thesection.\arabic{equation}}
The proof is established by showing that the total energy value (\ref{ee2min2}a) is non-increasing when sequence ($ {\pmb q},  {\pmb t}$) is updated.
According to the IPCTA-NOMA algorithm, we have
\begin{equation}\label{convergenceProofIPCTA}
U_{\text {Obj}}^{(v-1)}%&&\!\!\!\!\!\!\!\!\!\!
=E_{\text {Tot}}({\pmb  q}^{(v-1)}, {\pmb t}^{(v-1)})
%\nonumber\\
%&&\!\!\!\!\!\!\!\!\!\!
\geq E_{\text {Tot}}({\pmb  q}^{(v)}, {\pmb t}^{(v)})
%\nonumber\\
%&&\!\!\!\!\!\!\!\!\!\!
=U_{\text{Obj}}^{(v)},
\end{equation}
where the inequality follows from that (${\pmb  q}^{(v)}, {\pmb t}^{(v)}$) is a suboptimally optimal solution of problem  (\ref{ee2min2}) with fixed sets $\mathcal S_{ij}^{(v)}$, while  ($ {\pmb  q}^{(v-1)}, {\pmb t}^{(v-1)}$) is the initial feasible solution of problem  (\ref{ee2min2tdma}) with fixed sets $\mathcal S_{ij}^{(v)}$.
Furthermore, the total energy value (\ref{ee2min2}a) is always non-negative.
Since the total energy value (\ref{ee2min2}a) is non-increasing in each iteration according to (\ref{convergenceProofIPCTA}) and the total energy value (\ref{ee2min2}a) is finitely lower-bounded, the IPCTA-NOMA algorithm must converge.

%The proof is established by showing that the total energy value (\ref{ee2min2tdma}a) is non-increasing when sequence ($\hat {\pmb p}, \hat {\pmb q}, \hat {\pmb t}$) is updated.
%According to the IPCTA algorithm, we have
%\begin{eqnarray}\label{convergenceProofIPCTA}
%U_{\text {Obj}}^{(v-1)}&&\!\!\!\!\!\!\!\!\!\!
%=E_{\text {Tot}}(\hat {\pmb p}^{(v-1)}, \hat {\pmb  q}^{(v-1)}, \hat {\pmb t}^{(v-1)})
%\nonumber\\
%&&\!\!\!\!\!\!\!\!\!\!
%\geq E_{\text {Tot}}(\hat {\pmb p}^{(v)}, \hat {\pmb  q}^{(v)}, \hat {\pmb t}^{(v)})
%%\nonumber\\
%%&&\!\!\!\!\!\!\!\!\!\!
%=U_{\text{Obj}}^{(v)},
%\end{eqnarray}
%where the inequality follows from that ($\hat {\pmb p}^{(v)}, \hat {\pmb  q}^{(v)}, \hat {\pmb t}^{(v)}$) is the globally optimal solution of convex problem  (\ref{ee2min2tdma}) with fixed sets $\mathcal S_{ij}^{(v)}$, while  ($\hat {\pmb p}^{(v-1)}, \hat {\pmb  q}^{(v-1)}, \hat {\pmb t}^{(v-1)}$) is a feasible solution of convex problem  (\ref{ee2min2tdma}) with fixed sets $\mathcal S_{ij}^{(v)}$.
%Furthermore, the total energy value (\ref{ee2min2tdma}a) is always non-negative.
%Since the total energy value (\ref{ee2min2tdma}a) is non-increasing in each iteration according to (\ref{convergenceProofIPCTA}) and the total energy value (\ref{ee2min2tdma}a) is finitely lower-bounded, the IPCTA algorithm must converge.

\section{Proof of Theorem 6}
\setcounter{equation}{0}
\renewcommand{\theequation}{\thesection.\arabic{equation}}
We first show that problem (\ref{ee2min1tdma}) can be equivalently transformed into problem (\ref{ee2min2tdma}).
We introduce a set of new non-negative variables:
\begin{equation}\label{Th5eq1}
\hat p_{j} =t_{j} p_{j}, \hat q_i=t_{M+i} q_i, \quad \forall i \in \mathcal N, j \in \mathcal J_i.
\end{equation}
Substituting (\ref{Th5eq1}) into problem (\ref{ee2min1tdma}), we show that problem (\ref{ee2min1tdma}) is equivalent to  problem (\ref{ee2min2tdma}).

Then, we show that problem (\ref{ee2min2tdma}) is a convex problem.
Since constraints (\ref{ee2min2tdma}e)-(\ref{ee2min2tdma}i) are all linear, we only need to check that objective function (\ref{ee2min2tdma}a) and constraints (\ref{ee2min2tdma}b)-(\ref{ee2min2tdma}d) are convex.
Due to the fact that $-\bar u(|h_{nj}|^2 \hat q_n)$ is convex w.r.t. $\hat q_n$ from (\ref{appenDeq2_2}),
$-t_{M+n} \bar u\left(\frac{|h_{nj}|^2 \hat q_n}{t_{M+n}}\right)$ is convex w.r.t. ($t_{M+n}, \hat q_n$) according to the property of perspective function \cite[Page~89]{boyd2004convex}.
Thus, objective function (\ref{ee2min2tdma}a) is convex.
%Since $B \log_2\left(1+\frac{|h_{ij}|^2 \hat p_j}{ \sigma^2 } \right)$ is concave w.r.t. $p_{j}$,
%$B t_{j}\log_2\left(1+\frac{|h_{ij}|^2 \hat p_j}{ \sigma^2 t_j} \right)$ is concave w.r.t. ($t_{j}, p_{j}$) according to the property of perspective function \cite[Page~89]{boyd2004convex}.
%Thus, constraints (\ref{ee2min2tdma}b) are convex.
Analogously, constraints (\ref{ee2min2tdma}b)-(\ref{ee2min2tdma}d) can also be shown convex.
As a result, problem (\ref{ee2min2tdma}) is a convex problem.

\bibliographystyle{IEEEtran}
\bibliography{IEEEabrv,MMM}

\end{document}